\documentclass[12pt]{article}
\usepackage{amssymb}
\usepackage{amsmath}

\input{tcilatex}

\begin{document}

\input epsf

\begin{flushright}
hep-th/0609186 \\ PUPT-2205 \\MCTP-06-26
\end{flushright}
\vskip 0.5cm

\begin{center}
{\Large \textbf{Instantons, supersymmetric vacua, and \\[0pt]
emergent geometries} }

{\Large \vspace{ 20mm} }

{\normalsize {Hai Lin } }

{\normalsize \vspace{10mm} }

{\normalsize \emph{\ Michigan Center for Theoretical Physics, University of
Michigan, \\[0pt]
Ann Arbor, MI 48109-1040, USA} }\\[0pt]
{\normalsize \emph{\ and} }\\[0pt]

{\normalsize \emph{\ Department of Physics, Princeton University, \\[0pt]
Princeton, NJ 08544, USA} }

{\normalsize \vspace{0.2cm} }
\end{center}

\vspace{10mm}

\begin{abstract}
\medskip We study instanton solutions and superpotentials for the large
number of vacua of the plane-wave matrix model and a 2+1 dimensional Super
Yang-Mills theory on $R\times S^2$ with sixteen supercharges. We get the
superpotential in the weak coupling limit from the gauge theory description.
We study the gravity description of these instantons. Perturbatively with
respect to a background, they are Euclidean branes wrapping cycles in the
dual gravity background. Moreover, the superpotential can be given by the
energy of the electric charge system characterizing each vacuum. These
charges are interpreted as the eigenvalues of matrices from a reduction for
the 1/8 BPS sector of the gauge theories. We also discuss qualitatively the
emergence of the extra spatial dimensions appeared on the gravity side.
\end{abstract}

\newpage

\section{Introduction}

\renewcommand{\theequation}{1.\arabic{equation}} \setcounter{equation}{0}

Many supersymmetric field theories have a large number of supersymmetric
vacua. In the context of AdS/CFT correspondence \cite{Maldacena:1997re},\cite%
{Gubser:1998bc},\cite{Witten:1998qj}, when these field theories have string
theory dual descriptions, these vacua are dual to supergravity backgrounds.
These background geometries are the saddle points where string perturbation
theory can be expanded around.

In this paper we study instantons interpolating between different
supersymmetric geometries corresponding to different vacua of the gauge
theories. We associate each BPS vacuum geometry with a value of a
superpotetial, which is the vacuum expectation value of a particular
operator on the gauge theory side. The superponetial can be calculated
independently from gauge theory and gravity side. The superpotential from
the gravity side reveals the quantum nature of the emergent geometry \cite%
{Berenstein:2005aa} (see also \cite{Corley:2001zk},\cite{Berenstein:2004kk},%
\cite{Lin:2004nb}) as a strong coupling effect of gauge theory. The
emergence of extra spatial dimensions, extra relative to the boundary gauge
theory, can be explained in terms of the emergence of the geometries from
matrices in these gauge theories.

The specific gauge theories we will study for these issues are the 0+1
dimensional plane-wave matrix model \cite{Berenstein:2002jq} and a 2+1
dimensional super-Yang-Mills theory (SYM) on $R\times S^{2}$ \cite%
{Maldacena:2002rb},\cite{5brane} with sixteen supercharges. We chose these
theories for study since the explicit formulations of these theories on
gauge theory side and their supergravity dual backgrounds on the gravity
side are explicitly available \cite{5brane},\cite{Lin:2004nb}. The
fluctuations around these backgrounds are dual to perturbations around the
vacua of these gauge theories. These theories can be viewed as reductions of
the 3+1 dimensional $\mathcal{N}=4$ SYM on $S^{3}$ and on the Hopf fiber of $%
S^{3}$ or $S^{3}/Z_{k}.$ As a result, the string spectrum around these vacua
share some similarities especially in the $SO(6)$ sector \cite{5brane}. In 
\cite{5brane}, the near-BPS string spectrum was studied and shows a general
strong-weak coupling problem.

In order to better understand the vacuum structures of these theories, we
study the effects of instantons interpolating between different vacua. On
the gauge theory side, we analyze instanton solutions and instanton actions
in the weak coupling regime. We find that it is convenient to define a
superpotential and the difference of the superpotential between two vacua
gives the instanton action. In the gravity side, we try to find the dual
description of these instantons. We find in a large class of situations,
they are described by Euclidean branes wrapping some cycles in the BPS
geometries. The condition for instanton solutions in these cases is
described by the condition of embeddings of these branes. We compute their
actions and these give the strong coupling results of the instanton actions.
These instantons interpolate between initial and final vacua which are very
close to each other.

To understand the instanton action corresponding to initial and final vacua
that are not close to each other, we need to understand the instanton beyond
the Euclidean brane approximation on a background. We need to know the
non-perturbative superpotential of each vacuum. Each vacuum in the strong
coupling regime is described by an electrostatic configuration with charges
distributing on the disks under an external potential. The charges of the
electrostatic system can be interpreted as the eigenvalues of the matrices
from the gauge theory side. We found that the superpotential for each vacuum
is given by the energy of the eigenvalue system. We found consistency and
agreement of the superpotentials between gauge theory and gravity side,
except that the gravity result is the strong coupling result while the
result from classical gauge theory action is in the weak coupling limit. As
the couplings increase the actions of these theories receive quantum
corrections. The gravity answer shows quantum repulsions of the eigenvalues,
which was absent in the weak coupling limit, a feature similar to the
Dijkgraaf-Vafa matrix model \cite{Dijkgraaf:2002dh}.

The non-perturbative superpotential is intimately related to the emergence
of the geometries. We then study the emergence of all the extra spatial
dimensions, relative to the boundary gauge theories. One of our methods is
to embed the sector of these vacuum geometries into a larger sector of 1/8
BPS geometries associated with three harmonic oscillators, similar to \cite%
{Berenstein:2005aa}. We also try to explain the emergence of the
electrostatic system and the forces between eigenvalues from gauge theory.
On the gravity side the forces are described by the differential equations
governing the gravity solutions. We speculate on other possible approaches.

The organization of this paper is as follows: In section 2, we study
instanton solutions and superpotentials of these vacua on gauge theory side
in the weak coupling regime. The instanton actions are given by the
differences of the superpotentials between initial and final vacua. In
section 3, we study descriptions of these instantons and superpotentials on
gravity side. These include Euclidean brane analysis in the cases when
initial and final vacua are close to each other, and then an electrostatic
description of the superpotential for general vacua. These are results in
strong coupling regime of the gauge theories. In section 4, we discuss the
emergence of the extra coordinates and the electrostatic system on gravity
side, which encodes the information of the non-perturbative superpotential in
the strong coupling regime. Finally in section 5, we conclude and discuss
some related issues.

\section{Instantons on the gauge theory side}

\renewcommand{\theequation}{2.\arabic{equation}} \setcounter{equation}{0}

\label{gauge}The $U(N)$ plane wave matrix model has a gauge field $A_{0},$
three $SO(3)$ scalars $X_{i},~i=1,2,3$ and six $SO(6)$ scalars $%
X_{a},~a=4,5,...,9$ and their fermionic partners. They are all $N\times N$
Hermitian matrices. The action is written in for example \cite%
{Berenstein:2002jq},\cite{Dasgupta:2002ru},\cite{Kim:2002if},\cite%
{Kim:2002zg},\cite{Dasgupta:2002hx}. Here we use the action such that it
depends only on a dimensionless coupling constant $g_{ym0}^{2}/m^{3}$, where 
$m$ is originally\ the mass of the $SO(6)$ scalars, and can be set as 1 for
convenience. We have rescaled all fields so that the masses of $SO(6)$
scalars, $SO(3)$ scalars and fermions are 1, 2 and 3/2 respectively. The
action takes the form 
\begin{equation}
S=\frac{1}{g_{ym0}^{2}/m^{3}}\int dt~\mathrm{tr}\left( \frac{1}{2}%
D_{0}X_{a}D_{0}X_{a}-\frac{1}{2}X_{a}X_{a}+\frac{1}{4}[X_{a},X_{b}]^{2}...%
\right)
\end{equation}%
Due to the Myers term, the mass term and the commutator terms of $SO(3)$
scalars, the classical vacua are fuzzy spheres parametrized by $J_{i},$ the $%
N$ dimensional representation of $SU(2)~([J_{i},J_{j}]=i\epsilon
_{ijk}J_{k})\ $,$~X_{i}=2J_{i},~X_{a}=0.$ There are a large number of vacua
corresponding to partitions of integer $N$. In particular, there is a
\textquotedblleft trivial" vacuum $X_{i}=0,~X_{a}=0,$ which were conjectured
from gauge theory analysis to be a single NS5 brane wrapping $S^{5}$ \cite%
{Maldacena:2002rb}. The gravity dual of this vacuum was studied in \cite%
{5brane} (see also \cite{Ling:2006up},\cite{Ebrahim:2005sb}), which is a
linear dilaton background. The vacuum is dual to IIA little string theory on 
$S^{5}\times R,$ providing further evidence for the identification of the
single fivebrane. From the point of view of the M theory on asymptotically
plane-wave spacetimes, these vacua are 1/2 BPS M2 and M5 branes preserving
half of the supersymmetries of the asympotic plane-wave geometry. The
gravity duals of these vacua were studied in \cite{5brane}.

There could be in principle tunneling between different fuzzy sphere vacua,
exchanging some amount of D0 branes, if we perturb the theory. From the 11
dimensional point of view, this process is the transferring of the
longitudinal momenta of two fuzzy membranes. One naturally wonders whether
there are instanton solutions in the plane wave matrix model interpolating
different vacua. In fact, Yee and Yi \cite{Yee:2003ge} found a class of
analytical instanton solutions interpolating between a fuzzy sphere vacuum $%
X_{i}=2J_{i}~$and a trivial vacuum $X_{i}=0$. There could be two approaches
to the instanton action in the weak coupling gauge theory side. One is to
study a bound in the Euclideanized action corresponding to the action of the
instanton which was done in \cite{Yee:2003ge}, and the other is to study the
superpotential for each vacuum. The instanton equation can be read from the
superpotential or by supersymmetry transformation conditions\footnote{%
For example, the supersymmetry transformation conditions were utilized by 
\cite{Park:2002cb} in classification of the BPS equations for the plane-wave
matrix model.}.

We notice that in the weak coupling theory, the instanton solutions merely
involve the three $SO(3)$ scalars $X_{i}$ and their superpartners, but not
the fields in the $SO(6)$ sector. The solutions for $SO(6)$ scalars are $%
X_a=0$ for these instantons in the weak coupling theory. We can thereby
write a \textquotedblleft superpotential" \cite{Witten:1981nf},\cite%
{Salomonson:1981ug} for this supersymmetric quantum mechanics, for each
fuzzy sphere vacuum: 
\begin{equation}
W=\mathrm{tr~}(X_{i}X_{i}+i\frac{1}{3}\epsilon _{ijk}X_{i}X_{j}X_{k})
\label{W_bmn}
\end{equation}%
It's easy to check that the lagrangian for $X_{i}$ is correctly produced,
and the action is bounded by the difference of the superpotentials between
initial and final vacua: 
\begin{eqnarray}
S_{inst} &=&\frac{1}{g_{0}^{2}}\int_{-\infty }^{+\infty }d\tau ~\mathrm{tr}%
\left[ \frac{1}{2}(D_{\tau }X_{i})^{2}+\frac{1}{2}(\partial _{X_{i}}W)^{2}%
\right] \\
&=&\frac{1}{g_{0}^{2}}\left[ \int_{-\infty }^{+\infty }d\tau ~\mathrm{tr~}%
\frac{1}{2}(D_{\tau }X_{i}+\partial _{X_{i}}W)^{2}-W|_{\tau =-\infty }^{\tau
=+\infty }\right]
\end{eqnarray}%
where we denote $g_{ym0}^{2}/m^{3}=g_{0}^{2},~\tau =it.$

The instanton equation is the same as setting the square term to zero, 
\begin{equation}
D_{\tau }X_{i}+\partial _{X_{i}}W=D_{\tau }X_{i}+2X_{i}+i\epsilon
_{ijk}X_{j}X_{k}=0
\end{equation}
The equation is the same as in \cite{Yee:2003ge} and a class of analytical
solutions were found \cite{Yee:2003ge}:%
\begin{equation}
X_{block}^{i}(\tau )=X_{n_{p}\times n_{p}}^{i}(\tau )=2J_{n_{p}\times
n_{p}}^{i}\frac{1}{1+e^{2(\tau -\tau _{0})}}
\end{equation}%
Here $X_{block}^{i}$ is a $n_{p}\times n_{p}~$block matrix in $X^{i},$ and
we used $A_{0}=0$ gauge. This solution corresponds to the transition where a
fuzzy spheres with size $n_{p}$ at $\tau =-\infty ~$gradually shrinks to
zero and turns to a trivial vacuum $X_{block}^{i}=0~$at $\tau =+\infty $.
Moreover, we can have many such blocks labeled by $p$. So the instanton
action is the difference of the superpotentials $\Delta W~$between two vacua 
$X^{i}(\pm \infty )=2J_{i}|_{\pm \infty }$, 
\begin{eqnarray}
S_{inst} &=&-\frac{1}{g_{0}^{2}}\Delta W=-\frac{1}{g_{0}^{2}}\sum_{i=1,2,3}%
\frac{4}{3}\left( \mathrm{tr~}J_{i}^{2}|_{+\infty }-\mathrm{tr~}%
J_{i}^{2}|_{-\infty }\right)   \label{Sinst} \\
&=&\sum_{n_{p}}\frac{1}{g_{0}^{2}}\frac{n_{p}(n_{p}^{2}-1)}{3}
\end{eqnarray}%
where we used the identity that the second Casimir invariant $%
\sum_{i}J_{i,_{n\times n}}^{2}=\frac{(n^{2}-1)}{4}I_{n\times n}.$

One can define the $W$ for the trivial vacuum $X_{i}=0~$as some constant $%
W_{0}~$as a reference point and then the superpotential for an arbitrary
fuzzy sphere vacua can be defined as $W_{0}$ plus the difference term in (%
\ref{Sinst}). $W_{0}$ will be set to zero for convenience, because we only
need to know the differences of $W$ for different vacua.

We can label each vacuum by $N_{2}^{(i)}$ copies of $N_{5}^{(i)}$
dimensional irreducible representations. These satisfy $%
\sum_{i}N_{2}^{(i)}N_{5}^{(i)}=N$ . In the large $N_{5}^{(i)}$ limit, the
instanton action will be simplified to 
\begin{equation}
S_{inst}=-\frac{1}{g_{0}^{2}}\sum_{i}\frac{1}{3}N_{2}^{(i)}N_{5}^{(i)3}|_{-%
\infty }^{+\infty }  \label{Sinst_D0N2N5}
\end{equation}

In the supersymmetric quantum mechanics, the instanton amplitude between two
vacua $a$ and $b$ is proportional to a factor given by the difference of two
superpotentials $e^{-S_{inst}/\hbar }~\sim ~e^{-\frac{1}{g_{0}^{2}}%
(Wa-W_{b})}$. We set $\hbar =1$ in the second expression and also later
discussions$.$ This factor in the amplitude is largely suppressed if $%
g_{0}^{2}$ is small and the change $\sum_{i}\frac{1}{3}%
N_{2}^{(i)}N_{5}^{(i)3}|_{-\infty }^{+\infty }$ is large. However, there are
fermionic zero modes around these instanton solutions, the path integral for
the tunneling amplitude is zero, due to that the integration $\underset{l}{%
\prod }\int d\zeta _{l}~$of the collective coordinates of fermionic zero
modes $\zeta _{l}~$in the path integral gives zero. The vacuum energies for
these vacua will not be corrected and they are exactly protected vacua
quantum mechanically. Only fermionic operators with enough fermion fields
which can cancel the integration of the fermion collective coordinates will
receive instanton corrections to their vevs (see also \cite{Yee:2003ge}). In
our case, the instantons preserve 8 supersymmetries, and there are at least
8 fermionic zero modes, corresponding to the massless Goldstone fermions of
the broken supersymmetries.\footnote{%
This is in contrast to the bosonic matrix models. If we had a bosonic model
with the same action of the $SO(3)$ scalars but with no fermions, then the
energies of each fuzzy sphere vacuum would get corrected by these instanton
effects. The vacuum tunneling in a similar bosonic matrix model was analyzed
in detail by \cite{Jatkar:2001uh}.}

\begin{figure}[tbh]
\begin{center}
\epsfxsize=3.5in\leavevmode\epsfbox{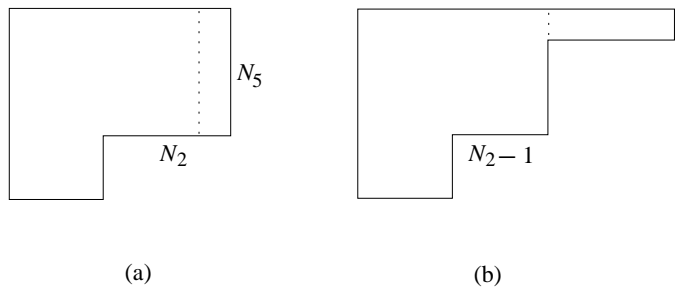}
\end{center}
\caption{Illustration of the condition for allowed instanton solutions in
the Young tableau representations: Comparing the initial Young tableau (a)
and the final Young tableau (b), there are only boxes moved from left to
right, but no boxes moved from right to left. In the example in the figure,
a column with $N_{5}$ boxes moved to the additional boxes on the right of
the first row. It corresponds to a instanton interpolating a fuzzy sphere
with size $N_{5}$ to a trivial representation.}
\label{young_tunnel}
\end{figure}

The method for solving general instanton solutions in plane-wave matrix
model and its moduli space has been discussed in detail by \cite%
{Bachas:2000dx},\cite{Yee:2003ge}. The solutions for bosonic and fermionic
zero modes around the instanton solutions have been analyzed by \cite%
{Yee:2003ge}. Both the instanton equation and their moduli space have
similarities with that of the domain wall solutions in $\mathcal{N}=1^{\ast }
$ SYM \cite{Polchinski:2000uf},\cite{Bachas:2000dx}. The condition for
allowed instanton solution between vacua $X^{i}(-\infty )~$and $%
X^{i}(+\infty )~$is determined by the initial and final representations of $%
SU(2)$ and the gauge group. Specifically, it requires that the final
representation modulo terms in the centralizer of it in the gauge group,
falls into the orbit of the initial representation under the gauge group
action, so that $gX^{i}(-\infty )g^{-1}$ and $X^{i}(+\infty )$ are
interpolated (fore more details see \cite{Bachas:2000dx}). In the
representation of Young tableaux, the condition for allowed instanton
solution is that when comparing the initial Young tableau and the final
Young tableau, there are some boxes moved from left to right, but there are
no box allowed to move from right to left. An example is illustrated in
figure \ref{young_tunnel}. The condition also implies that any fuzzy sphere
vacuum can be interpolated with the trivial vacuum. 

Now we turn to the 2+1 dimensional super-Yang-Mills on $S^{2}\times R$ $,$
which is a limit from the plane wave matrix model. The 2+1 dimensional SYM
on $S^{2}\times R$ we discuss here is the theory when we reduce the $%
\mathcal{N}=4$ SYM on $S^{3}\times R$ on the Hopf fiber of the $S^{3}~$\cite%
{5brane}$.$ The gauge fields in the $\mathcal{N}=4$ SYM reduce to the gauge
fields in the 2+1 d SYM plus an additional scalar $\Phi .$ The $SO(6)$
scalars $X_{a},~a=4,5,...,9$ reduce to their components invariant on the $%
S^{1}.$ There are totally seven scalars. The gravity duals of its vacua were
studied in \cite{5brane}. The consistent truncation of Kaluza-Klein modes
from $\mathcal{N}=4$ SYM to this theory was studied in detail by \cite%
{Ishiki:2006rt} \footnote{%
The consistent truncation from $\mathcal{N}=4$ SYM to plane-wave matrix
model were studied in for example \cite{ Kim:2003rz},\cite{Klose:2003qc},%
\cite{Fischbacher:2004iu},\cite{5brane}, where a phenomenon of the
redefinition of the gauge coupling constant was observed, in both weak and
strong coupling regimes. See also related earlier work in \cite%
{Okuyama:2002zn}.}.

The theory can also be obtained from expanding plane-wave matrix model
around its fuzzy sphere vacuum, so under the continuum limit (e.g. \cite%
{Kabat:1997im}) of the matrix regularization of the large fuzzy sphere, it
becomes a theory on $S^{2}\times R$. Suppose we start from the vacuum with $%
N_{2}$ fuzzy spheres, each of size $\overline{N}_{5}\gg 1,~$we have $N_{2}%
\overline{N}_{5}=N,$ where $N$ is the number of original D0 branes. We
expand around this vacuum. Fluctuations of$~X_{i}$ about $2J_{i}$ ($%
\overline{N}_{5}\times \overline{N}_{5})~$are decomposed into gauge fields
on the fuzzy sphere and a scalar $\Phi ~$\cite{Maldacena:2002rb}$.$ The
scalar $\Phi $ describes the fluctuations of the sizes of the fuzzy spheres
around $\overline{N}_{5}.$

Due to its relation with the plane-wave matrix model, the vacua and their
superpotentials analyzed above in the plane wave matrix model carry over to
the $U(N_{2})$ 2+1 d SYM. For the $U(2)$ theory, the vacua are quite simple,
which are characterized by one integer $n,$ this is the eigenvalue of $\Phi
\sim $\textrm{diag}$(n,-n)$. The tunneling issue was briefly discussed in 
\cite{Maldacena:2002rb}. It was further studied in detail by \cite%
{Lee:2005dd}, who presented the instanton equation and solved a class of
analytical solutions for the $U(2)$ theory. We will analyze the general $%
U(N_{2})$ case, for we already know the superpontial for each vacuum via the
relation to plane-wave matrix model. One can map the instanton solutions in
plane-wave matrix model to those of the 2+1 d SYM theory.

The action of the 2+1 d SYM can be written in two ways. One is in terms of
scalar $\Phi $ and gauge field $A.$ Another is in terms of three scalars $%
Y_{i}$ which are combinations of the $\Phi $ and gauge fields $A$%
\begin{equation}
Y^{i}=e^{i}\Phi +\epsilon ^{ijk}e^{j}A_{k},~~A_{i}=\epsilon
^{ijk}Y_{j}e^{k},~\ \ \Phi =e^{i}Y^{i}
\end{equation}%
where $e_{i}$ is a unit vector in $R^{3},~e_{i}e_{i}=1.~$The formalism in
terms $\Phi $ and $A$ is more direct from a gauge theory point of view. The $%
Y_{i}$ originates in the plane-wave matrix model, from the fluctuations of
the $SO(3)$ scalars around the fuzzy sphere $Y_{i}=X_{i}-2J_{i}.~$The
formalism in terms of $Y_{i}$ is more convenient for seeing its origin from
the plane-wave matrix model.

We will write the action as in eqn (2.9) in \cite{5brane}, further, we write
it in such a way that it depends only on a dimensionless coupling constant $%
g_{ym2}^{2}/m$, where $m$ was originally\ the mass of $SO(6)$ scalars and
can be set to 1 for convenience. We rescale all fields so that the masses of 
$SO(6)$ scalars, $\Phi $ and fermions are 1, 2 and 3/2 respectively. The
action takes the form 
\begin{equation}
S=\frac{1}{g_{ym2}^{2}/m}\int dt\frac{1}{4}d^{2}\Omega ~\mathrm{tr}\left( 
\frac{1}{2}D_{0}X_{a}D_{0}X_{a}-\frac{1}{2}X_{a}X_{a}+\frac{1}{4}%
[X_{a},X_{b}]^{2}...\right)
\end{equation}%
This is the same in effect as we set the radius of $S^{2},~1/\mu =1/2~$in 
\cite{5brane}.

Because the vacua satisfy an equation $f+2\Phi =0,$ where $f$ is the $S^{2}$
component of the gauge field strength and is quantized into integers time a
half, the vacuum is characterized by the eigenvalues of $\Phi $ 
\begin{equation}
\Phi =\frac{1}{4}~\mathrm{diag}(n_{5}^{(1)},n_{5}^{(2)},...,n_{5}^{(N_{2})})~
\end{equation}%
after an unitary transformation \cite{5brane}$.~$Notice that $%
n_{5}^{(i)}=N_{5}^{(i)}-\overline{N}_{5}~(n_{5}\ll \overline{N}_{5})$ are
the fluctuations of the sizes of $N_{2}~$fuzzy spheres.

We want to use the superpotential in the plane-wave matrix model to define
the one in this theory. Some of the mappings of fields between the two
theories are 
\begin{equation}
~Y_{i}=X_{i}-2J_{i},\,~~\mathrm{tr}_{\overline{N}_{5}\times \overline{N}%
_{5}}\rightarrow \overline{N}_{5}\int d^{2}\Omega ,~~\ \ [J_{j},]\rightarrow
L_{j,~~~}\ L_{j}=-i\epsilon _{ijk}e_{k}\partial _{i}~
\end{equation}%
where $e_{i}$ is a unit vector in $R^{3},~e_{i}e_{i}=1.$ The superpotential
can be derived from the one in plane-wave matrix model (\ref{W_bmn}) as 
\begin{equation}
W=\overline{N}_{5}\int d^{2}\Omega ~\mathrm{tr}_{N_{2}\times
N_{2}}(Y_{i}Y_{i}+i\frac{1}{3}\epsilon _{ijk}Y_{i}Y_{j}Y_{k}+i\epsilon
_{ijk}Y_{i}L_{j}Y_{k})\ +W_{J_{i}}
\end{equation}%
where the second term is a large constant term 
\begin{equation}
W_{J_{i}}=\sum_{i=1,2,3}\frac{4}{3}~\mathrm{tr}_{N\times N}J_{i}^{2}
\label{W_fuzzysphere}
\end{equation}%
which corresponds to the superpotential for the vacuum with exact $N_{2}$
numbers of fuzzy spheres of size $\overline{N}_{5},$ and the first term is
the difference of the superpoential with respect to the vacuum $\Phi =%
\mathrm{diag}(0,0,...,0).~$The instanton equation is%
\begin{equation}
D_{\tau }Y_{i}+\partial _{Y_{i}}W=D_{\tau }Y_{i}+2Y_{i}+i\epsilon
_{ijk}Y_{j}Y_{k}+2i\epsilon _{ijk}Y_{i}L_{j}Y_{k}=0
\end{equation}%
The solutions to these equations in principle can be lifted from those
solutions in the plane-wave matrix model for $X_{i},$ so the moduli space of
these solutions are in principle known.

We want to have also the expression in the variables of $\Phi $ and gauge
field $A$. As discussed in appendix of \cite{5brane}, that the 2+1 d SYM on $%
R\times S^{2}$ can be lifted to a 3+1 d SYM on $R^{2}\times S^{2},$ which
can also be obtained from reduction of a 4+1 d SYM on $R^{2}\times S^{3}$ on
the Hopf fiber of $S^{3}.$ Interestingly, \cite{Lee:2005dd} noticed that the
instanton equation in the Euclideanized 2+1 d SYM on $R\times S^{2}$ is the
same as the self-dual Yang-Mills equation on $R^{2}\times S^{2}~$\cite%
{Lee:2005dd}$.$ If we lift the Euclideanized 2+1 d SYM on an $x_{3}$
direction, we can define the four dimensional gauge field strength $\mathcal{%
F}$ on $R^{2}\times S^{2}$%
\begin{equation}
\mathcal{F}=fd^{2}\Omega +(-2\Phi )d\tau dx_{3},~~\ast \mathcal{F}=(-2\Phi
)d^{2}\Omega +fd\tau dx_{3}
\end{equation}%
Note that the self-dual equation $\mathcal{F}\mathcal{=}\ast \mathcal{F}$ is
equivalent to the equation of motion $f+2\Phi =0.~$The part of the action
that does not involve the $SO(6)$ scalars and is inherited from the $SO(3)$
sector in the plane-wave matrix model can be written as a $\frac{1}{2}\int 
\mathcal{F\wedge }\ast \mathcal{F}$ term.$~$So the action is bounded by 
\begin{eqnarray}
S_{E} &=&\frac{1}{2}\int \mathcal{F\wedge }\ast \mathcal{F=}\frac{1}{4}\int (%
\mathcal{F-\ast \mathcal{F)}\wedge (}\ast \mathcal{F-F)+}\frac{1}{2}\int 
\mathcal{F\wedge F} \\
&\geqslant &\frac{1}{2}\int \mathcal{F\wedge F}\sim \int d^{2}\Omega ~%
\mathrm{tr~}\Phi ^{2}\text{ \ \ }\qquad
\end{eqnarray}%
The bound is satisfied only when $\mathcal{F}=\ast \mathcal{F}$. Thus after
matching parameters, we get that the instanton action should be 
\begin{equation}
S_{inst}=-\frac{1}{g_{0}^{2}}\Delta W=-\frac{1}{g_{0}^{2}}16N_{2}\overline{N}%
_{5}(\mathrm{tr~}\Phi ^{2}|_{+\infty }-\mathrm{tr~}\Phi ^{2}|_{-\infty })=-%
\frac{1}{g_{0}^{2}}N_{2}\overline{N}_{5}\sum_{i}n_{5}^{(i)2}|_{-\infty
}^{+\infty }  \label{Sinst_D2}
\end{equation}%
This expression is of course consistent with the $U(2)$ case studied by \cite%
{Maldacena:2002rb},\cite{Lee:2005dd}.

From the 2+1 d SYM point of view, $\overline{N}_{5}$ is absorbed into the
definition of the coupling constant via the relation $\frac{1}{g_{ym2}^{2}}=%
\frac{\mu ^{2}\overline{N}_{5}}{g_{ym0}^{2}}~$\cite{Maldacena:2002rb}. So we
have the instanton action for the 2+1 d SYM%
\begin{equation}
S_{inst}=-\frac{1}{g_{2}^{2}}4N_{2}(\mathrm{tr~}\Phi ^{2}|_{+\infty }-%
\mathrm{tr~}\Phi ^{2}|_{-\infty })=-\frac{1}{4g_{2}^{2}}N_{2}%
\sum_{i}n_{5}^{(i)2}|_{-\infty }^{+\infty }
\end{equation}%
where $g_{2}^{2}=g_{ym2}^{2}/m,$ purely in terms of parameters in the 2+1 d
SYM.

\section{Gravity description}

\label{gravity}

\renewcommand{\theequation}{3.\arabic{equation}} \setcounter{equation}{0}

\subsection{Euclidean branes}

\label{ebrane} In this section we turn to the gravity analysis of the
instantons and superpotentials for the vacua of the two theories. In some
regimes, the instanton solutions we studied in the previous section can be
approximated as Euclidean branes wrapping some cycles in the dual
backgrounds \cite{5brane}.

The backgrounds dual to the vacua of these theories have an $SO(3)$ and $%
SO(6)$ symmetry and thereby contain an $S^{2}~$and$~S^{5}$ factors, and the
remaining three coordinates are time, $\rho $ and $\eta ,$ where $\rho $ is
a radial coordinate. These backgrounds are regular because the dual theories
have mass gaps. The gravity equations of motion reduce to a three
dimensional Laplace equation for $V$, it is in the space of $\rho $,$~\eta $
if we combine $\rho $ with an $S^{1}~$\cite{Lin:2004nb}. The regularity
condition requires that the location where the $S^{2}$ shrinks are disks (or
lines if without the $S^{1}$) at constant $\eta _{i}$ in the $\rho $,$~\eta $
space, while the location where the $S^{5}$ shrinks are the segment of the $%
\rho =0$ line between nearby two disks \cite{5brane}. $V$ is regarded as a
electric potential and there are charges on the disks. Due to flux
quantizations, the charge and $\eta $ distance are quantized. The charges
are distributed in such a way that the disks are equipotential surfaces, and
the charge densities on the edge of the disks vanish. The cycle we will
embed the Euclidean D2-brane is the 3-cycle at $\rho =0$ between two disks,
where the $\rho =0$ line combine the $S^{2}$ to form a 3-cycle.

\begin{figure}[tbh]
\begin{center}
\epsfxsize=3.5in\leavevmode\epsfbox{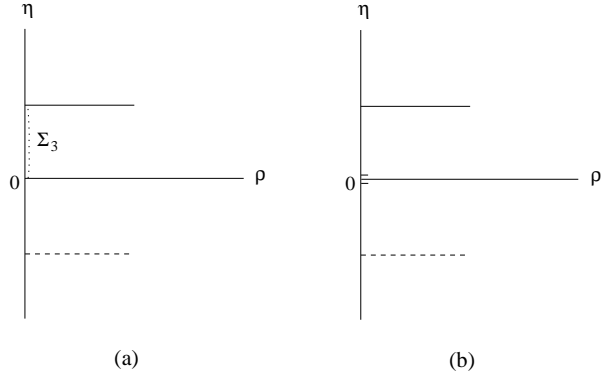}
\end{center}
\caption{The configurations before and after the transition. In (a) there is
a large disk with $N_{2}~$units of charge above the infinite disk. In (b)
the large disk has $N_{2}-1~$units of charge and there is another small disk
near origin with $N_{5}$ units of charge. There is a $\Sigma _{3}$ cycle
that is along $\protect\rho =0$ and between the disks, formed together with
the $S^{2}$. Below the infinite disk, there are image disks.}
\label{tunneling}
\end{figure}

Since we use an Euclidean brane in a geometric background, such analysis is
only an approximation when the transition is very slight or perturbative in
nature. The geometries dual to two different vacua are different, so a
non-perturbative analysis is generically needed, and it will be discussed in
next section. The geometries before and after transition are approximated as
the same background in this section. The wrapping of the brane should
satisfy DBI-WZ action and global conditions.

We first look at some regimes which obviously can be compared from both
sides. We first look at the plane-wave matrix model. We look at in the gauge
theory side transition from a vacuum with $N_{2}$ copies of fuzzy spheres
with size $N_{5},$ to the vacuum with $N_{2}-1~$copies of fuzzy spheres with
size $N_{5}$ plus $N_{5}$ copies of size 1 (or a single NS5 brane with $%
N_{5}~$D0 branes)$.$ We keep $N_{5}$ moderately large and $N_{2}$ large. In
the dual gravity description, each fuzzy sphere is a disk with charge
proportional to $N_{2},$ and located at a distance $\eta _{0}=\frac{\pi }{2}%
N_{5}$ above the $\eta =0$ plane (for more details, see section 2.2 of \cite%
{5brane}). In the gravity picture, it corresponds to transition from a large
disk with charge $Q=\frac{\pi ^{2}}{8}N_{2}~$above the $\eta =0$ plane, to a
large disk with charge $Q^{\prime }=\frac{\pi ^{2}}{8}(N_{2}-1)$ plus a very
small disk near the origin (see figure \ref{tunneling}). We consider the
small disk and its image as a small dipole with dipole moment $2(Q-Q^{\prime
})\eta _{0}=\frac{\pi ^{3}}{8}N_{5}~$near the origin, and that they do not
back-react to the geometry. The total dipole moment, which corresponds to $N$
D0 branes, is conserved.

There is a nontrivial 3-cycle at $\rho =0,$ between the large disk and the $%
\eta =0$ plane. The cycle has topology of an $S^{3},$ with the $S^{2}$
shrinks only at $\eta =0$ and $\eta =\eta _{0}.$ The Euclidean D2 brane
wraps the $S^{2}$ and its Euclidean time $\tau $ is embedded along $\eta $
direction extending from $\eta =0$ to $\eta =\eta _{0}.$ We should find $%
\eta (\tau )$ as a function of $\tau $ from the DBI-WZ action. Such
Euclidean branes preserve half of the total 16 supercharges, since they
extend along $\eta$ and have an additional projection condition for the
Killing spinors. This Euclidean D2-brane has $N_5$ Euclidean D0-brane
charges, and it mediates the transferring of one unit of D2 charge and $N_5$
units of D0 charges, see figure \ref{tunneling}. We utilize the general
background solution in eqn (2.20-2.24) of \cite{5brane}. It is parametrized
by an electric potential $V$ as a function of $\rho ,\eta$.

The DBI part of the action is 
\begin{eqnarray}
S_{DBI}/\tau _{2} &=&-\int d\tau d^{2}\Omega e^{-\Phi }\sqrt{\det
G_{\parallel }}\sqrt{\det G_{\perp }+\det (2\pi \alpha ^{\prime }F-B)} \\
&=&-\int d\tau 4\pi e^{-\Phi }\sqrt{(-g_{tt})+g_{\eta \eta }(\partial _{\tau
}\eta )^{2}}\sqrt{g_{22}^{2}+(\pi N_{5}\alpha ^{\prime }-2\eta -\frac{2%
\overset{\cdot }{V}\overset{\cdot }{V^{\prime }}}{\Delta })^{2}}
\end{eqnarray}%
where $\tau _{2}~$is the Euclidean D2 brane tension,~$g_{tt},g_{\eta \eta
},g_{22}$ are various metric components on the $t$, $\eta ,~$and$~S^{2}~$%
directions$~$in the background \cite{5brane}. $G_{\parallel },$ $G_{\perp }~$%
are the time component and the spherical components of the pull-back metric,
and $\det G_{\perp }=g_{22}^{2}\sin ^{2}\theta ,$ $2\pi \alpha ^{\prime
}F-B=(\pi N_{5}\alpha ^{\prime }-2\eta -\frac{2\overset{\cdot }{V}\overset{%
\cdot }{V^{\prime }}}{\Delta })\sin \theta d\theta d\phi $ according to flux
quantization. We have chosen the gauge that $2\pi \alpha ^{\prime }F-B$ is
zero where $\eta $ is at the top disk. Dot and prime are the derivatives
w.r.t. $\log \rho $ and $\eta $. The WZ part of the action is 
\begin{eqnarray}
S_{WZ}/\tau _{2} &=&-\int [C_{3}+(2\pi \alpha ^{\prime }F-B)\wedge C_{1}] \\
&=&-\int d\tau 4\pi \frac{\overset{.}{4V}}{\overset{..}{V}-\overset{.}{2V}}[-%
\overset{.}{V}-\overset{.}{V}^{\prime }(\frac{1}{2}\pi N_{5}\alpha ^{\prime
}-\eta )]
\end{eqnarray}

Due to the background $H_{3}~$flux, the Euclidean D2 brane worldvolume gauge
field has a source of $N_{5}~$units of magnetic charges, and these are
precisely cancelled by the $N_{5}~$D0 branes ending on it. This is
consistent for wrapping the brane on this cycle. This is similar to the
situation of Euclidean D2 branes wrapping on the $SU(2)$ group manifold with 
$H_{3}~$flux discussed in \cite{Maldacena:2001xj}.

The expression of the action can be simplified if we expand the electric
potential near $\rho \approx 0,$ as 
\begin{equation}
V=\sum_{i=0,even}^{\infty }(-1)^{\frac{i}{2}+1}\frac{K_{0}^{(i)}}{\left(
i!!\right) ^{2}}\rho ^{i}=-K_{0}+\frac{K}{4}\rho ^{2}-\frac{K^{\prime \prime
}}{64}\rho ^{4}+...,~\ ~~~~~~~K\equiv K_{0}^{\prime \prime }
\end{equation}%
where $K$ is only a function of $\eta$. $K$ has some properties such as $%
K>0,~\frac{-8K}{K^{\prime \prime }}>0,$ in order to satisfy the regularity
condition for the geometry. The action is then 
\begin{eqnarray}
\frac{S_{E}}{4\pi \tau _{2}} &=&-\int d\tau \left( \frac{-8K}{K^{\prime
\prime }}\right) ^{\frac{1}{2}}\sqrt{[K+K^{\prime }(\frac{\pi }{2}N_{5}-\eta
)]^{2}-\frac{1}{2}KK^{\prime \prime }(\frac{\pi }{2}N_{5}-\eta )^{2}}\sqrt{%
\frac{-8K}{K^{\prime \prime }}+(\partial _{\tau }\eta )^{2}}  \notag \\
&&+\int d\tau \frac{-8K}{K^{\prime \prime }}[K+K^{\prime }(\frac{\pi }{2}%
N_{5}-\eta )]
\end{eqnarray}%
We have set $\alpha^{\prime}=1$. Since the action does not explicitly depend
on $\tau ,$ we can have a conserved quantity $H=P\partial _{\tau }\eta -L=%
\mathrm{const.}=E,$ where $P=\partial _{\partial _{\tau }\eta }L.$ It is
obvious that the constant solution $\eta =\frac{\pi }{2}N_{5}~$is a special
case of a class of solutions to the equations, and from this solution we get 
$E=0.~$Setting $E=0 $, we get the equation of motion 
\begin{equation}
\partial _{\tau }\eta =-\frac{2K(\frac{\pi }{2}N_{5}-\eta )}{K+K^{\prime }(%
\frac{\pi }{2}N_{5}-\eta )}
\end{equation}%
The solution for the Euclidean D2 brane is 
\begin{equation}
e^{-2\tau }=\frac{K}{(\frac{\pi }{2}N_{5}-\eta )}
\end{equation}%
There could be an overall constant on the RHS, but it can be absorbed by
shifting $\tau$.

We plug in the solution back into the action and integrate out to get the
final answer of the Euclidean D2-brane action 
\begin{eqnarray}
S_{E} &=&-4\pi \tau _{2}\int_{0}^{\frac{\pi }{2}N_{5}}2K(\frac{\pi }{2}%
N_{5}-\eta )d\eta \\
&=&-\frac{2}{\pi }[V(\eta _{0})-V(0)-\eta _{0}V^{\prime }(0)]
\label{Sinst_gravity}
\end{eqnarray}%
where we used integration by parts several times to arrive at the last
expression, and we used $\tau _{2}=\frac{1}{4\pi ^{2}},~\alpha ^{\prime
}=1,~\eta _{0}=\frac{\pi }{2}N_{5}.$ $V(\eta )$ in the expression is
understood as the potential $V(\rho ,\eta )$ along the $\rho =0~$line$.$ The
expression (\ref{Sinst_gravity}) has the right property of the invariance
under the shift of $V$ by a linear term in $\eta ,$ since the gravity
solution should be invariant under this shift.

To get more intuition, let's look at the example of two infinite disks,
which is the NS5 brane soultion in \cite{5brane}. We wrap a Euclidean D2
brane on the throat region of the NS5 brane solution. For that solution \cite%
{5brane}, 
\begin{equation}
V=\frac{1}{\widetilde{g}_{0}}I_{0}(r)\sin \theta ,~~~~r=\frac{2\rho }{N_{5}}%
,~\ ~~\theta =\frac{2\eta }{N_{5}}
\end{equation}%
where $\widetilde{g}_{0}$ is a constant, so $K=-V^{\prime \prime }|_{\rho
=0}={\frac{4}{N_{5}^{2}}}\widetilde{g}_{0}^{-1}\sin \theta .~$The action of
the Euclidean D2 brane is 
\begin{equation}
\frac{S_{E}}{8\pi \widetilde{g}_{0}^{-1}\tau _{2}}=-\int d\tau \sqrt{(\pi
-\theta )^{2}(c^{2}+1)+4sc(\pi -\theta )+2s^{2}}\sqrt{8+(\partial _{\tau
}\theta )^{2}}+\int d\tau 4[s+c(\pi -\theta )]
\end{equation}%
where $c=\cos\theta, s=\sin\theta$, $\theta \in \lbrack 0,\pi ],~$and the
equation of motion is 
\begin{equation}
\partial _{\tau }\theta =\frac{2(\pi -\theta )s}{s+c(\pi -\theta )}
\end{equation}%
The solution for the Euclidean D2-brane is then 
\begin{equation}
e^{-2\tau }=e^{-2\tau _{0}}\frac{\sin (\pi -\theta )}{\pi -\theta }
\end{equation}%
where $\tau _{0}$ is a constant. When $\theta =\pi $, $\tau (\pi )=\tau
_{0},~$when$~\theta =0$, $\tau (0)=+\infty $, so this transition corresponds
to transferring a small charge (corresponding to one unit of $N_{2}$)$~$from
the top disk to the bottom disk. The Euclidean D2 brane action for this
solution is $S_{E}=\frac{2}{\widetilde{g}_{0}}.$

Of course, the solution in the form (\ref{Sinst_gravity}) works for general
configurations of disks in the plane-wave matrix model, when there are
nearby two disks. The Euclidean D2 branes wrapping on that cycle mediate
transferring of charges between two disks while keeping the total dipole
moment fixed. It also works for the 2+1 d SYM. In that case, the Euclidean
D2 brane mediates similar transferring of charges between two nearby disks,
while the total dipole moment of the system is kept to zero, because the
theory is expanded from the plane-wave matrix model and the disks are
already in the center of charge frame.

There are issues of the range of validity that need to be addressed. One is
that the calculation of the embedding of the Euclidean D-brane needs that
the gravity backgrounds are weakly curved, this requires that the two disks
where the cycle is in between are quite large, and their separation is also
quite large. This needs the parameters $(g_{ym0}^{2}/m^{3})N_{2},~N_{5}$ to
be relatively large. On the other hand, $N_{5}~$had better not be too large,
this is because from the Euclidean D2 brane point of view, there are $N_{5}$
D0 branes ending on its worldvolume, if $N_{5}$ is too large we would
consider these $N_{5}$ D0 branes as a single NS5 brane, and the brane
configuration should be changed. From the disk point of view the small disk
near the origin should be much smaller then the top disk, so $N_{5}\ll
N_{2}. $ And these conditions have already guaranteed that the charges on
the Euclidean D2 brane are much smaller than the background charges.

A second issue of the range of validity is that the Euclidean brane
calculations only describe the transition between very close vacua or very
close backgrounds. The geometric backgrounds before and after the transition
are approximated as the same geometry. So in principle we need
non-perturbative description which could interpolate between very different
backgrounds, as long as it is allowed by the condition in section \ref{gauge}%
. This will be discussed more in section \ref{electric energy} in terms of
the non-perturbative superpotentials.

The condition for the instanton solution as discussed in section \ref{gauge}
in the gauge theory side is that there are boxes moving from left to right
only, in the Young tableaux. The number of boxes are the number of D0 branes
transferred. This condition in the gravity side in our case is the
transferring of charges from the higher disk to the lower disk. The process
in figure \ref{tunneling} corresponds to the illustration in figure \ref%
{young_tunnel}.

The instantons and Euclidean branes we studied so far describe tunnelings
between spherical giant gravitons and point giant gravitons in the 11d
plane-wave geometry. This can be related to the limit of those instantons
tunneling between spherical giant gravitons and point giant gravitons in $%
AdS_{4,7}\times S^{7,4}$ studied by \cite{Hashimoto:2000zp}, \cite%
{Grisaru:2000zn}, (see also \cite{Das:2000fu}). The difference is that the
giant gravitons (\cite{McGreevy:2000cw},\cite{Hashimoto:2000zp},\cite%
{Grisaru:2000zn}) in their analysis of the tunneling are probe branes in the 
$AdS_{4,7}\times S^{7,4}$ geometries and in our case we have replaced the
giant gravitons with the back-reacted geometries taking into account the
back-reaction of them on the plane-wave geometry. It is analyzed both in 
\cite{Hashimoto:2000zp} and \cite{Yee:2003ge} that these instantons have 8
supersymmetries, and are 1/4 BPS objects from the point of view of the
asymptotic geometries. In our case, the kappa symmetry condition on the
Euclidean D2 brane will impose an additional projection condition, with
respect to the one for the background\footnote{%
In this case, we have the kappa symmetry projector on the Euclidean D2
brane: $\Gamma \epsilon =(\sqrt{1+(\frac{\mathcal{F}_{22}}{g_{22}})^{2}}%
i\Gamma _{\overline{\tau }}\Gamma _{\overline{\theta }}\Gamma _{\overline{%
\phi }}-\frac{\mathcal{F}_{22}}{g_{22}}\Gamma _{\overline{\theta }}\Gamma _{%
\overline{\phi }}\Gamma _{11})\epsilon =\epsilon .$~Here $\mathcal{F}%
_{22},g_{22}$ are the $S^{2}~$components of the worldvolume 2-form flux and
the pull-back metric, $\Gamma _{\overline{\tau }}=E_{\tau }^{\overline{\tau }%
}\Gamma _{i}\frac{\partial x^{i}}{\partial \tau }=(\sqrt{\frac{g_{\eta \eta }%
}{-g_{tt}}}\partial _{\tau }\eta \Gamma _{\overline{\eta }}+i\Gamma _{%
\overline{t}})/\sqrt{1+\frac{g_{\eta \eta }}{-g_{tt}}(\partial _{\tau }\eta
)^{2}},$ $\epsilon =S(x^{i})\epsilon _{0},$ where $\epsilon _{0}$ is a
constant spinor subject to projection conditions $\Gamma _{\overline{t}%
}\Gamma _{11}\epsilon _{0}=i\Gamma _{\overline{\eta }}\Gamma _{\overline{%
\theta }}\Gamma _{\overline{\phi }}\epsilon _{0}=\epsilon _{0}.$}. Again,
there will be at least 8 fermionic zero modes corresponding to the broken
supersymmetries and the fermionic zero modes suppress the tunneling
amplitude and forbid the mixing of the vacua.

One can also consider an Euclidean NS5 brane wrapping a six-cycle enclosing
the disk, which is the $\Sigma _{6}$ cycle in \cite{5brane}. This Euclidean
NS5 brane has $N_{2}~$units of Euclidean D0 charges. It interpolates between
a vacuum of $N_{5}$ NS5 branes with size $N_{2}~$and the vacuum with $%
N_{5}-1 $ NS5 branes with size $N_{2},~$plus $N_{2}~$D0 branes. On the
gravity side, the transition corresponds to lowering the disk by one unit of
the height. This instanton solution is allowed from gauge theory side but
its action is zero in the weak coupling limit. From the gravity side and
strong coupling point of view, the action should be non-zero. The existence
of this Euclidean NS5 brane is a quantum effect from gauge theory point of
view. Since the upper surface and lower surface of the disk are treated
differently, it is more convenient to use the coordinates of the $%
x_{1}=x_{11},~x_{2}=V^{\prime },~y=\overset{.}{V}$ appeared in the Toda
equation \cite{Lin:2004nb},\cite{5brane}. From 11d point of view, it's a
Euclidean M5 brane wrapping $S^{5},$ with its worldvolume $\tau ~$extending
along $x_{2}~$of the M5 strip at $y=0$ and traveling along the $x_{1}$
circle, with $N_{2}~$units of momenta $P_{x_{1}}$ conjugate to $x_{1}$. From
the point of view of the IIA Euclidean NS5 brane, there is a worldvolume
scalar corresponding to $x_{1},$ and the action \cite{Bandos:2000az} can be
written as\footnote{%
Note if the NS5 brane with $N_{2}$ D0 charges wraps $T^{5}~$instead of $%
S^{5},$ the configuration can be T-dualized to $(1,N_{2})$ 5-brane in type
IIB theory after taking T dualities five times \cite{Papadopoulos:1997je}.
The IIA NS5 brane action with D2 or D4 charges have been studied in detail
in for example \cite{Bena:2000fz},\cite{Bena:2000va}.} 
\begin{eqnarray}
S_{E}/\tau _{5} &=&-\int d\tau d^{5}\Omega e^{-2\Phi }\sqrt{%
(-g_{tt})+g_{x_{2}x_{2}}(\partial _{\tau }x_{2})^{2}+e^{2\Phi }\mathcal{F}%
_{\tau }^{2}}\sqrt{\det g_{55}}+\int [B_{6}+C_{5}\wedge \mathcal{F}]  \notag
\\
\mathcal{F} &=&(\partial _{\tau }x_{1}-C_{t})d\tau
\end{eqnarray}%
where $\tau _{5}=\frac{1}{(2\pi )^{5}},\alpha ^{\prime
}=1,~g_{tt},~g_{x_{2}x_{2}},~g_{55}$ are the components of the metric along
the $t,$ $x_{2},~S^{5}~$directions, $B_{6},C_{5}$ are the pull-backs of the
potentials dual to $B_{2},C_{3},~$and$~\mathcal{F}$ is the worldvolume
one-form flux. We look at solutions that $P_{x_{1}}=\frac{\partial L_{\tau }%
}{\partial (\partial _{\tau }x_{1})}=\frac{N_{2}}{R_{11}},$ that is
quantized into integers $N_{2}~$corresponding to D0 charges$,$ and then the
equation of motion will yield $x_{2}(\tau )$ as a function of $\tau ,$ which
characterizes the embedding of the brane. If we expand the solution near an
isolated disk, we can approximate it by the solution of a single disk, which
was the solution (2.49)-(2.54) of \cite{5brane} corresponding to the 2+1 d
SYM. Using this solution, we then estimate the action to be $S_{E}\sim
g_{ym2}^{-4}(g_{ym2}^{2}N_{2})^{4/3}\sim g_{ym0}^{-4}\eta _{0}^{2}\rho
_{0}^{4},$ where $\eta _{0}~$and $\rho _{0}~$are the height and size of the
disk, and $g_{ym0}=g_{0}$ is the coupling constant in the plane-wave matrix
model. This is a quantum effect and strong coupling result which is not seen
in the weak-coupling gauge theory.

\subsection{Electric charge system and superpotential}

\label{electric energy}

Now we analyze what the quantity $W$ is in the electrostatic picture. In the
gauge theory, the instanton action is proportional to the difference of the
superpotentials of the two vacua as defined by $S_{inst}=-\frac{1}{g_{0}^{2}}%
\Delta W~$in (\ref{Sinst}).$\ $In the case of the two vacua that is very
similar to each other, we used the approximation of an Euclidean brane
action to match the instanton action. Of course, the Euclidean brane action
will be the strong-coupling result. For the configuration of two vacua
studied in section \ref{ebrane}, involving the exchange of a unit of charge
(corresponding to a unit of $N_{2}$) between two disks, we arrived at the
expression 
\begin{equation}
S_{E}=-\frac{2}{\pi }[V(\eta _{0})-V(0)-\eta _{0}V^{\prime }(0)]
\end{equation}%
This quantity is proportional to the change of the total energy $U$ in the
electrostatic system (including image charges). Note that the relation
between D2 brane number and electric charge is $Q=\frac{\pi ^{2}}{8}N_{2}~$%
\cite{5brane}, so the charges are in units of $\frac{\pi ^{2}}{8}$,~the
charge transfer in this process is $\Delta Q=\frac{\pi ^{2}}{8}.$ The energy
that the unit charge and its image charge released is 
\begin{equation}
-\Delta U_{\mathrm{ch}\arg \mathrm{e}}=2\Delta Q[V(\eta _{0})-V(0)]
\end{equation}%
On the other hand, the total dipole of the charge system is conserved, the
dipole had been lost by the amount $2\Delta Q\eta _{0}$ by transferring the
unit charge from the top disk. After the transferring process, we have the
small disk (and also its image disk) near the origin, as in figure \ref%
{tunneling}(b). We make an approximation when a pair of small disks near the
origin is approximated as a point. We thereby have a small dipole with
dipole moment $2\Delta Q\eta _{0}$ at the origin. The electric field at the
origin is $-V^{\prime }(0),$ which is pointing upward, so the coupling
between the dipole and the electric field there gives an negative energy,
this is the energy the dipole released: 
\begin{equation}
-\Delta U_{\mathrm{dipole}}=(2\Delta Q\eta _{0})[-V^{\prime }(0)]
\end{equation}%
So the total energy difference of the system between final and initial
configuration is 
\begin{equation}
\Delta U_{\mathrm{total}}=-2\Delta Q[V(\eta _{0})-V(0)-\eta _{0}V^{\prime
}(0)]
\end{equation}%
which is just proportional to the Euclidean D2 brane action, in appropriate
unit. Of course this definition of the change of the total energy is
invariant under the shift of the potential $V~~$by a linear term$~b\eta .$
The energy difference will not be changed due to this shift of $V$. This is
because the charge $\Delta Q,$ will release an additional energy $2\Delta
Qb\eta _{0},$ on the other hand, due to that there is an additional electric
field downward with magnitude $b,$ the dipole's energy is increased by $%
2\Delta Q\eta _{0}b,$ which exactly cancels the additional energy released
from the charge. So this expression is consistent with the definition of the
change of the total energy.

We therefore can identify (in the strong coupling regime) $S_{inst}=S_{E}=%
\frac{8}{\pi ^{3}}\Delta U,$ and thereby the superpotential for each vacuum
can be defined in terms of the energy of the electric configuration
corresponding to that vacuum 
\begin{equation}
U=-\frac{\pi ^{3}}{8g_{0}^{2}}W
\end{equation}%
up to an overall constant shift which is not important when comparing
differences. The minus sign is due to our conventions. So the superpotential
of the system is proportional to the total energy of charges (including
image charges)%
\begin{equation}
W=-\frac{16g_{0}^{2}}{\pi ^{3}}\sum_{i}Q_{i}V_{i}  \label{W_nonperturb}
\end{equation}%
and $W_{0}$ which corresponds to the single NS5 brane vacuum is set to zero
for convenience. $i$ represents different groups of charges at $\eta =\eta
_{i}~$with potential $V_{i}.$ Now we see that the factor of the instanton
amplitude is given by the energy difference between the two configurations
of the electric charge system or the eigenvalue system. The factor in the
instanton amplitude is largely suppressed if the energy difference is large.
Note that there are still other parameters like the $\alpha ^{\prime }$ and $%
\hbar ,$ which have been set to 1 for convenience. These charges are
associated with eigenvalues from the $SO(6)$ scalars and will be discussed
in section \ref{emergence}.

It might be useful to notice that the expressions (\ref{W_nonperturb}) can
be reduced to integrals of forms in $\rho ,\eta $ space. We define one forms
locally $dV=\partial _{\eta }Vd\eta +\partial _{\rho }Vd\rho ,~dq=2\pi \rho
\ast _{2}dV=2\pi (-\partial _{\eta }V\rho d\rho +\partial _{\rho }V\rho
d\eta ),~$where $\ast _{2}$ is the flat space epsilon symbol in $\rho ,\eta $
space. The Laplace equation for $V$ is just $d(dq)=0.$

We choose one-cycle $\Pi _{i}~$enclosing each disk, and one-cycle$~\Gamma
_{i}$ between each disk and a reference disk (whose potential is set as
zero). We see that 
\begin{equation}
Q_{i}=\int_{\Pi _{i}}dq,~~~~\ ~~V_{i}=\int_{\Gamma _{i}}dV
\end{equation}%
Since $\Pi _{i}$ cycle and the $S^{5}$ forms a 6-cycle, and $\Gamma _{i}$
cycle and $S^{2}$ forms a 3-cycle in 9 spatial dimensions, the $\Pi _{i}$
and $\Gamma _{i}$ cycles do not intersect,$~\Pi _{i}\cap \Gamma _{i}=0.$ We
can thereby form a two-cycle by the cup product of the two cycles, so (\ref%
{W_nonperturb}) can be written as 
\begin{equation}
W=-\frac{16g_{0}^{2}}{\pi ^{3}}\int_{\Pi _{i}\cup \Gamma _{i}}dq\wedge dV=-%
\frac{16g_{0}^{2}}{\pi ^{3}}\int_{\Pi _{i}\cup \Gamma _{i}}|\nabla
V|^{2}2\pi \rho d\rho \wedge d\eta
\end{equation}%
where $\nabla V$ is the gradient of $V$ in $\rho ,\eta $ space.\ So we have
defined the superpotential of a boundary gauge theory by purely forms
appeared in the bulk of gravity duals. This view is very similar in spirit
to the definition of the non-perturbative superpotential in 3+1 d $\mathcal{N%
}=1$ gauge theories via a flux background after geometric transition, e.g. \cite{Cachazo:2001jy}. 

Let's compare the superpotential in the weak coupling case. The charges in
the electric system originate from the eigenvalues of the matrices in the $%
SO(6)$ sector in the gauge theories. We will discuss this more in section %
\ref{emergence}. Classically, these eigenvalues do not interact with each
other. They distribute under the external electric potential, which are
quadratic ($\sim \rho ^{2}$) in the gravity dual of either theories \cite%
{5brane}. So they are sitting on top of each other at fixed position $\eta
_{i}$ at $\rho =0$. Quantum mechanically, the eigenvalues repel each other
and they are not coincident. They thus form extended disks. The classical
and quantum pictures, correspond to weak coupling and strong coupling
effects for the superpotential. In the weak coupling analysis, the
superpotential does not involve any dependence on the $SO(6)$ scalars, so
this must get corrections when coupling increases and becomes large\footnote{%
Some proposals for the quantum mechanically corrected action of the
plane-wave matrix model taking into account the quantum effects of the $%
SO(6) $ fields have been put forward, for example, in \cite{Nastase:2004te},%
\cite{Lozano:2005kf} using fuzzy five-sphere constructions. It is also
possible to use the approach of \cite{Taylor:1999gq}.}.

For the plane-wave matrix model, the background potential is $V_{b}=\frac{8}{%
g_{s}}(\rho ^{2}\eta -\frac{2}{3}\eta ^{3}),$ where $g_{s}=4\pi
^{2}g_{ym0}^{2}$ (e.g. \cite{Itzhaki:1998dd}) and $g_{s}$ is the string
coupling that appears at the asymptotic D0 brane near horizon geometry \cite%
{Lin:2004kw}. Since in the gravity side, $m$ is the unit of energy for $%
\frac{|g_{tt}|}{g_{55}}$ \cite{5brane}, which is 1 in our case, so we have $%
m=1.~$In the weak-coupling regime of the gauge theory, since the charges sit
at $\rho =0~$and$~\eta =\eta _{i},$ they feel potential $V_{i}=-\frac{4}{%
3\pi ^{2}g_{0}^{2}}\eta _{i}^{3},$ so the superpotential from (\ref%
{W_nonperturb}) is: 
\begin{equation}
W=\frac{64}{3\pi ^{5}}\sum_{i}Q_{i}\eta _{i}^{3}=\sum_{i}\frac{1}{3}%
N_{2}^{(i)}N_{5}^{(i)3}
\end{equation}%
where we used$~N_{2}^{(i)}=\frac{8Q_{i}}{\pi ^{2}},~N_{5}^{(i)}=\frac{2\eta
_{i}}{\pi }~$\cite{5brane} from flux quantization conditions. This is the
same as the field theory calculation in weak-coupling regime, cf. (\ref%
{Sinst_D0N2N5}). $\ $The factor of the instanton amplitude is largely
suppressed if $g_{s}$ is small and the change of the energy of the system is
large. 
\begin{figure}[tbh]
\begin{center}
\epsfxsize=3.0in\leavevmode\epsfbox{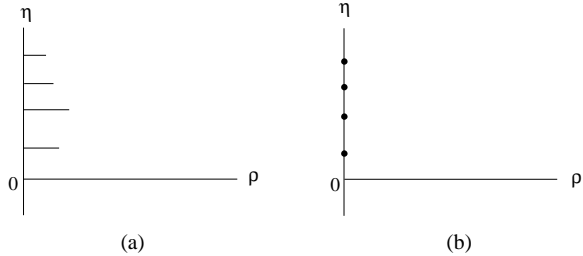}
\end{center}
\caption{Comparison of strong coupling and weak coupling superpotentials. In
(a) the charges or eigenvalues are extended into disks due to eigenvalue
interactions in strong coupling regime. In (b) they are on top of each other
at $\protect\rho =0~$line and not extended in the $SO(6)$ directions in the
weak coupling limit.~The superpotential is given by the energy of the
electric charge system. }
\label{weakcoupling}
\end{figure}

The vacua of the 2+1 d SYM are inherited from that of plane-wave matrix
model. In the disk configuration of the plane-wave matrix model, if we look
at the geometry locally near a group of disks that is isolated from all
other disks, including their images, then the geometry locally approaches to
the solution dual to 2+1 d SYM theory. The vacuum of the 2+1 d SYM thereby
is also characterized by the total energy of the electrostatic system. A
special case is that when we start from plane-wave matrix model and look at
the configuration where all the disks are very high above $\eta =0$ plane
and they are relatively near each other. Their average position is $\eta
_{0}=\frac{\pi }{2}\overline{N}_{5}$. We define new coordinate relative to
the average position $\widetilde{\eta }=\eta -\eta _{0}=\frac{\pi }{2}%
n_{5},~(\widetilde{\eta }\ll \ \eta _{0})~$and $n_{5}=N_{5}-\overline{N}%
_{5},~(n_{5}\ll \overline{N}_{5})$. The total number of the D2 brane charges
is $N_{2}=\sum_{i=disks}N_{2}^{(i)}.~$So in the new coordinate frame, \ this
is the configuration dual to the $U(N_{2})$ 2+1 d SYM. $n_{5\text{ }}$is
proportional to the location of each disk in the new frame. The case with
all $n_{5}=0$ corresponds to a single disk at $\widetilde{\eta }=0,$ and is
the simplest vacuum with unbroken $U(N_{2}).$The number $\overline{N}_{5}$
is used in the matrix regularization for the $S^{2}.$ The electric potential
near this group of disks is%
\begin{equation}
V=\frac{2}{\pi ^{2}g_{0}^{2}}(\rho ^{2}\eta -\frac{2}{3}\eta ^{3})=\frac{%
2\eta _{0}}{\pi ^{2}g_{0}^{2}}(\rho ^{2}-2\widetilde{\eta }^{2})-\frac{4}{%
3\pi ^{2}g_{0}^{2}}\eta _{0}^{3}+...
\end{equation}%
They basically feel the same external potential $\frac{2\eta _{0}}{\pi
^{2}g_{0}^{2}}(\rho ^{2}-2\widetilde{\eta }^{2})~$as in the gravity dual of
the 2+1 d SYM theory discussed in \cite{5brane}, with a large constant shift
due to our expansion. The dot terms are the linear term in $\widetilde{\eta }
$ and higher order terms in $\widetilde{\eta },$ which are not important.~

In the weak coupling regime, the charges are not extended in the $\rho $
direction. The total energy of the system (excluding the large constant
term) gives the superpotential from the relation (\ref{W_nonperturb})%
\begin{equation}
W=-\frac{16g_{0}^{2}}{\pi ^{3}}\sum_{i}Q_{i}~\frac{2\eta _{0}}{\pi
^{2}g_{0}^{2}}(-2\widetilde{\eta }_{i}^{2})=N_{2}\overline{N}%
_{5}\sum_{i}n_{5}^{(i)2}
\end{equation}%
The large constant term from the $\eta _{0}^{3}$ term in the electric
potential, corresponds to $W_{J_{i}}=\sum_{i}\frac{4}{3}~\mathrm{tr}%
_{N\times N}J_{i}^{2}$ term for the fuzzy sphere we are expanding around in
the plane-wave matrix model, cf. (\ref{W_fuzzysphere}).

We thereby find that the weak-coupling superpotential analyzed in section %
\ref{gauge}, which does not involve $SO(6)~$scalars, is the situation in the
gravity side when the eigenvalues are not extended in the $\rho $ direction,
or the $SO(6)$ directions. This is when the eigenvalues do not have
interactions between each other. To bring in the interactions, we need
quantum effect, and this is shown in the strong coupling regime in the disk
picture. These are illustrated in figure \ref{weakcoupling}.

Now we briefly discuss some technical aspects of solving the value of the
superpotential in the strong-coupling regime, for the disk configurations.
We first discuss the situation with a single disk above $\eta =0$ plane (\ref%
{Sinst_gravity}), and then the situation with some arbitrary numbers of
disks (\ref{W_nonperturb}).

We first discuss the situation of a pair of positively and negatively
charged disks under the background potential $V_{b}=\frac{8}{g_{s}}(\rho
^{2}\eta -\frac{2}{3}\eta ^{3}).$ The size of the disks is $\rho _{0}$ and
their separation is $2\eta _{0}.$ Apart from the background, the charges
have mutual repulsions from other charges on the disk and attractions from
the image disk. Suppose the charge density and the potential on the disk is $%
\sigma (\rho )$ and $V_{0}.$ We can scale out everything and the problem
then is only determined by the ratio $\kappa =\frac{\eta _{0}}{\rho _{0}}.$
We make the rescaling$~y=\eta /\rho _{0},~x=\rho /\rho _{0},~\ \sigma (\rho
)=\frac{8}{g_{s}}\rho _{0}^{2}g(x),~V=\frac{8}{g_{s}}\rho _{0}^{3}v.$

The integral equation in these variables is (which involves dimensionless
functions only) 
\begin{equation}
\{(x^{2}\kappa -\frac{2}{3}\kappa ^{3})+\int_{0}^{2\pi }\int_{0}^{1}[\frac{%
g(x^{\prime })x^{\prime }dx^{\prime }d\phi }{\sqrt{x^{2}-2xx^{\prime }\cos
\phi +x^{\prime 2}}}-\frac{g(x^{\prime })x^{\prime }dx^{\prime }d\phi }{%
\sqrt{x^{2}-2xx^{\prime }\cos \phi +x^{\prime 2}+4\kappa ^{2}}}]\}_{\forall
x\in \lbrack 0,1]}=\mathrm{const.}  \label{inteqn_g}
\end{equation}%
with the supplementary condition that $g(1)=0$ due to vanishing of the
charge at the edge of the disk. When $g(x)$ is solved, it is also a function
of $\kappa$.

The integral equation is very difficult to solve due to its complicated
kernel. Provided we know the function $g(x),$ since we already know the
total charge of the disk, what we need to compute is the potential $V$ on
the disk. Since the disk is an equipotential surface, we can know it from
the potential at $\rho =0,$ this consists of a contribution from the
background and a contribution from the charges. The value of the
superpotential (\ref{W_nonperturb}) is 
\begin{equation}
W/(\frac{1}{3}N_{2}N_{5}^{3})=1-\frac{2}{3}\kappa ^{-3}2\pi
\int_{0}^{1}dx~g(x)\left[ 1-\frac{x}{(x^{2}+4\kappa ^{2})^{1/2}}\right]
\end{equation}%
It contains two terms. The first term \textquotedblleft 1" on the RHS gives
a contribution to the superpotential that is the same as the weak coupling
gauge theory result $W=\frac{1}{3}N_{2}N_{5}^{3}$, it's the result if the
charges were all sitting at $\rho =0.~$The second contribution is from the
integral term, resulted from the effect of mutual interactions among
charges. We do not assume that $\kappa $ has to be small, so in principle,
if we were looking for perturbative corrections to the weak coupling answer,
the integral term gives the corrections relative to the first piece, in the
large $\kappa ~$regime. The $\kappa $ depends on $N_{2}$ in a very
non-linear way. In the small $\kappa $ limit, it is estimated in \cite%
{5brane} that $\rho _{0}^{4}=\frac{1}{2}\pi ^{4}g_{ym0}^{2}N_{2},~\eta _{0}=%
\frac{\pi }{2}N_{5},$ so $\kappa =\frac{\pi }{2}N_{5}(\frac{1}{2}\pi
^{4}g_{ym0}^{2}N_{2})^{-1/4}$.

The problem of solving the integral equation has been simplified by Abel
transforming the kernel in (\ref{inteqn_g}) as done in \cite{Ling:2006up}.
The new integral equation is 
\begin{equation}
f(x)-\int_{-1}^{1}{\frac{1}{\pi }}{\frac{2\kappa }{4\kappa ^{2}+(x-t)^{2}}}%
f(t)dt=1-2\alpha x^{2}  \label{inteq}
\end{equation}%
where $f(t)$ is the Abel transform of $g(x)$, $~f(t)={\frac{2\pi }{\beta }}%
\int_{t}^{1}{\frac{rg(r)dr}{(r^{2}-t^{2})^{\frac{1}{2}}},}$ $\beta =v_{0}+%
\frac{2}{3}\kappa ^{3},$ with the additional condition $f(1)=0,~f^{\prime
}(1)$ is bounded \cite{Ling:2006up}, and $\alpha $ is a constant determined
by these boundary conditions. The new kernel is the one appears in the
Love's integral equation \cite{love},\cite{kp},\cite{hutson}. The solution
in this form seems very promising. The kernel can even be simplified a
little bit more in the small $\kappa $ limit \cite{kp}.

Now let's discuss the calculation of the superpotential for the
electrostatic configuration with an arbitrary number of disks. Once we
specify the charges $Q_{i}~$of the disks and their heights $\eta _{i}$, what
we need to know is the electric potential of each disk $V_{i},$ in order to
compute the superpotential (\ref{W_nonperturb}). We can thereby use the
conformal transform techniques in \cite{5brane}.

In \cite{5brane}, the complex variables were introduced: $w=2\partial
_{z}V=\partial _{\rho }V-i\partial _{\eta }V~$and$~z=\rho -\rho _{0}+i\eta .$
In the large disk limit, the general solution for the plane-wave matrix
model is 
\begin{equation}
\partial _{w}z=\prod\limits_{j=1}^{n}\frac{(w-ia_{j})}{(w-ic_{j})}(-ia_{n+1})
\end{equation}%
where there are $n$ finite disks located at $ia_{j},$ and $n$ fivebrane
throats located at $ic_{j},~$and finally the infinite disk at $ia_{n+1}.$ We
can integrate it to get the potential of each disk 
\begin{equation}
V_{i}=\func{Re}\int wdz=-\func{Re}\int_{ia_{i}}^{ia_{n+1}}\prod%
\limits_{j=1}^{n}\frac{(w-ia_{j})}{(w-ic_{j})}(-ia_{n+1})wdw
\end{equation}%
and then plug this in (\ref{W_nonperturb}). The rest is to figure out the
relation between $a_{i},c_{i}$ and $N_{2}^{(i)},N_{5}^{(i)}~$\cite{5brane}.

\section{Emergence of gravity picture and eigenvalue system}

\label{emergence}

\renewcommand{\theequation}{4.\arabic{equation}} \setcounter{equation}{0}

In the previous sections, we find that the instanton action is associated
with the superpotential of each vacuum, which is in turn related to the
emergence of the geometries. The geometries are emergent from matrices in
these theories. There are different types of emergence in our cases. The
emergence of odd dimensional spheres and the even dimensional spheres seem
to be different in some cases. In this section, we will discuss
qualitatively the emergence of the extra coordinates and the electrostatic
picture.

For the plane-wave matrix model, we have to explain the emergence of 9
directions, $S^{2},~S^{5}$, $\eta ~$and $\rho $. We first discuss the
emergence of $S^{2}$ in the plane-wave matrix model. This is by the $SU(2)$
commutation relations of $J_{i}~$(e.g. \cite{Kabat:1997im}). Other higher
even dimensional spheres may similarly emerge in theories constructed using
higher even dimensional fuzzy spheres for example \cite{Castelino:1997rv},%
\cite{Ramgoolam:2001zx}.

These fuzzy two-spheres have radial motions, which are characterized by the
scalar $\Phi ,$ most conveniently seen in the 2+1 d SYM. Comparing sections %
\ref{gauge} and \ref{gravity}, we find that the eigenvalues of $\Phi $ are
proportional to $n_{5}^{(1)},n_{5}^{(2)},...,n_{5}^{(N_{2})},$ which map to
the heights of each disk in $\eta $ direction:$~\eta _{1},\eta _{2},...,\eta
_{N_{2}}.$~So the $\eta $ direction is emerged from the eigenvalues of $\Phi 
$.

Next we come to the emergence of the odd dimensional sphere $S^{5}~$and the
direction $\rho .$ From intuition, due to the reduction from $\mathcal{N}=4$
SYM, the $SO(6)$ sector in these theories are quite similar to $\mathcal{N}%
=4 $ SYM. So the emergence of $S^{5}$ should be similar to that of $\mathcal{%
N}=4$ SYM \cite{Berenstein:2005aa}.

In order to address this problem, we embed these vacuum geometries into the
larger sector of 1/8 BPS geometries. These 1/8 BPS geometries are dual to
1/8 BPS states in these gauge theories. We can look at 1/8 BPS states
satisfying $E-(J_{A}+J_{B}+J_{C})=0,$ where $E$ is the energy and $%
J_{A},J_{B},J_{C}$ are the three $U(1)$ $R$-charges associated with the
excitations by three complex scalars from the $SO(6)$ scalars. These 1/8 BPS states are in turn embedded in an even
larger Hilbert space. Finally our BPS vacuum geometries are the ground
states in the 1/8 BPS sector.

Though quite similar, a difference of the theories we analyzed, relative to $%
\mathcal{N}=4$ SYM on $S^{3}\times R,$ is that we have many ground state
geometries characterized by a broken gauge symmetry $U(N)\rightarrow
\prod\limits_{i}U(N^{(i)}).$ The most convenient theory is the 2+1 d SYM
where a particular vacuum is specified by the eigenvalues of the scalar $%
\Phi .$ If there are $N^{(i)}$ coincident eigenvalues, then there is a $%
U(N^{(i)})$ symmetry. These eigenvalues form an extended disk in the
electric charge picture, but this could have alternative descriptions if we
embed the geometries into larger sectors. If the disks are far away from
each other, the repulsion of charges on each disk is quite similar to the
case of a single disk. In this limit, we study only the case of a single
disk, that is, we look at the block of the $SO(6)$ scalars corresponding to $%
N^{(i)}$ and only excite $U(N^{(i)})$ invariant 1/8 BPS states. Then finally we can excite all the blocks and build $%
\prod\limits_{i}U(N^{(i)})$ invariant 1/8 BPS states.

For our purposes, we can consider first a single disk vacuum in the 2+1 d
SYM with unbroken $U(N)~$symmetry, and then other cases will be quite
similar. These 1/8 BPS states can be described by the harmonic
oscillators of the $SO(6)$ scalars and only s-wave modes of the $S^{2}$ are
relevant for this class of BPS excitations. We use the methods similar to
those of \cite{Berenstein:2005aa},\cite{Corley:2001zk},\cite%
{Berenstein:2004kk}\footnote{%
One can also use the approach of collective fields for the three complex
matrices, for example in \cite{deMelloKoch:2002nq}, although in the loop
variables the other two matrices are treated as impurities with respect to
the first matrix, the method is in principle general.}. The action that is
relevant for these excitations of $SO(6)$ scalars is the matrix quantum
mechanics%
\begin{equation}
S=\frac{\pi }{g_{ym2}^{2}/m}\int dt~\mathrm{tr}\left( \frac{1}{2}\overset{%
\cdot }{X_{a}}\overset{\cdot }{X_{a}}-\frac{1}{2}X_{a}X_{a}+\frac{1}{4}%
[X_{a},X_{b}]^{2}\right) 
\end{equation}%
where $a,b=4,5,...,9$ and we have set $\mu =2$ in the lagrangian. We set $%
A_{0}=0$ and have a Gauss law constraint on $U(N)~$gauge singlet physical
states, $\frac{\delta L}{\delta A_{0}}\left\vert \Psi \right\rangle =0.~$%
Furthermore, the commutator terms\footnote{%
Strings can also emerge on the backgrounds of these emergent geometries,
when we turn on the commutator terms. They are fluctuations of the impurity
matrices on the background of other matrices (corresponding to the
background geometry), see the approaches of, for example, \cite%
{Berenstein:2005jq},\cite{Donos:2005vm},\cite{Rodrigues:2005ec},\cite%
{Vazquez:2006hd} from gauge theory side.} will not be included in this class
of BPS excitations because they contribute positively to the energy but not
the $R$-charges. Then we define the conjugate momenta $P_{a}~$for each $%
X_{a}.~$The energy and $R$-charges are 
\begin{eqnarray}
E &=&~\frac{1}{2}\mathrm{tr~}(P_{a}P_{a}+X_{a}X_{a}),~~~J_{A}=\mathrm{tr~}%
(P_{4}X_{5}-P_{5}X_{4}),~ \\
J_{B} &=&\mathrm{tr~}(P_{6}X_{7}-P_{7}X_{6}),~~~J_{C}=\mathrm{tr~}%
(P_{8}X_{9}-P_{9}X_{8})
\end{eqnarray}%
We can form the usual three complex scalars $Z_{L},$ and their conjugate
momenta $\Pi _{L}=-i\frac{\partial }{\partial Z_{L}^{\dagger }},~(L=A,B,C).$
They satisfy the standard commutation relations$~[Z_{L}^{mn},\Pi
_{L}^{\dagger m^{\prime }n^{\prime }}]=i\hbar \delta _{nm^{\prime }}\delta
_{n^{\prime }m},~[Z_{L},\Pi _{L}]=0,$ so we can define creation/annihilation
operators 
\begin{equation}
a_{L}^{\dagger }=\frac{1}{\sqrt{2}}(Z_{L}^{\dagger }-i\Pi _{L}^{\dagger
}),~~\ \ b_{L}^{\dagger }=\frac{1}{\sqrt{2}}(Z_{L}^{\dagger }+i\Pi
_{L}^{\dagger })~
\end{equation}%
and their conjugates. The energy and $R$-charge operators are then written
as 
\begin{equation}
E=\mathrm{tr}\sum_{L}(a_{L}^{\dagger }a_{L}+b_{L}^{\dagger }b_{L}),~~~J_{L}=%
\mathrm{tr~}(a_{L}^{\dagger }a_{L}-b_{L}^{\dagger }b_{L})
\end{equation}%
So it's clear in order to get 1/8 BPS states, we should keep only the three $%
a_{L}^{\dagger }$ oscillators but not the three $b_{L}^{\dagger }$
oscillators. We have then in this 1/8 BPS subspace of the Hilbert space a
three dimensional harmonic oscillator Hilbert space. Its phase space is six
dimensional. The singlet condition tells us that we should look at states of
products of traces%
\begin{equation}
\left\vert \Psi \right\rangle =\prod\limits_{j=1}^{k}\mathrm{tr~}%
(a_{A}^{\dagger })^{l_{j}}(a_{B}^{\dagger })^{m_{j}}(a_{C}^{\dagger
})^{n_{j}}\left\vert 0\right\rangle ~  \label{1-8BPSWAVE}
\end{equation}%
with $l_{j},m_{j},n_{j}\leqslant N$. Because the commutator terms vanish, by
unitary transformations we may simultaneously put $Z_{A},Z_{B},Z_{C}$ into
the form of triangular matrices, each of which is a sum of a diagonal matrix
and an off-diagonal triangular matrix. We integrate out the unitary matrices
and also the off-diagonal triangular matrices\footnote{%
We refer the readers to \cite{Takayama:2005yq},\cite{Yoneya:2005si} for more
detailed discussions and more subtleties for the case of one holomorphic
complex matrix.}. Let the diagonal matrix elements be $z_{Ai},z_{Bi},z_{Ci}.~
$The wave function corresponding to (\ref{1-8BPSWAVE}) is of the form%
\begin{equation}
\Psi \sim \Delta (z_{Ai},z_{Bi},z_{Ci})\prod\limits_{j=1}^{k}\mathrm{~}%
(\sum_{i}z_{Ai}{}^{l_{j}})(\sum_{i}z_{Bi}{}^{m_{j}})(%
\sum_{i}z_{Ci}{}^{n_{j}})e^{-\frac{1}{2}\sum_{i}z_{Ai}{}\overline{z}%
_{Ai}{}+z_{Bi}{}{}\overline{z}_{Bi}{}+z_{Ci}\overline{z}_{Ci}{}}~
\end{equation}%
up to some normalization factors, including the factor from integrating out
off-diagonal matrices. $\Delta (z_{Ai},z_{Bi},z_{Ci})$ is a Van-de-monde
determinant from integrating out unitary matrices. See similar wave
functions in \cite{Berenstein:2005aa}, and its 1/2 BPS case \cite%
{Corley:2001zk},\cite{Berenstein:2004kk},\cite{Caldarelli:2004ig},\cite%
{Takayama:2005yq},\cite{Yoneya:2005si}. There is an interesting property.
The wave function, besides the universal Gaussian factor, is a holomorphic
function in $z_{Ai},z_{Bi},z_{Ci},$ due to the 1/8 BPS condition. This
holomorphicity provides some simplifications in the wave functions. Wave
functions that depends also on ${}\overline{z}_{Ai}{},{}{}\overline{z}%
_{Bi}{},{}\overline{z}_{\func{Ci}}$ are non-BPS. The complex structure is
related to the symplectic structure in the phase space $C^{3}.$ Each $z_{Li}~
$coordinates in the phase space are paired by a physical real coordinate and
its conjugate momentum, $z_{Li}=x_{Li}+ip_{Li}$ \cite{Berenstein:2005aa}.

In the large $N$ limit, the wave function has a thermodynamic
interpretation, the wave function squared describes the probability of a
particular eigenvalue distribution in phase space \cite{Berenstein:2005aa}. 
The saddle point approximation of the wave function gives a droplet
configuration in the phase space. The polynomial part of the wave function
provides repulsion forces between eigenvalues. The Gaussian factor gives a
spherically symmetric quadratic potential. The ground state should be a
spherically symmetric droplet with $SO(6)$ symmetry in $C^{3}$, which is
bounded by an $S^{5}.$ Because of the spherical symmetry of the ground state
droplet, the $S^{5}$ emerges in the gravity description. We see that the
eigenvalues are extended in the radial direction of $R^{6},~$the $\rho $
coordinate in the gravity description is really mapped from the radial
direction of the phase space $R^{6}.$ In the electric charge picture,
charges are extended along $\rho ,$ if we tensor the $\rho $ with the $S^{5}$%
, the charges are really distributing in a six dimensional space. The
electric disks emerge from the eigenvalues of the $SO(6)$ matrices. This
would be more clear if we had all the regular 1/8 BPS geometries of these
theories. The $S^{5}$ then will be a particular enhanced symmetry for only
the ground state geometries.

If we start from 1/4 BPS sector or 1/2 BPS sector, we will have similar but
reduced droplet descriptions. They are described by two or one dimensional
harmonic oscillators similar to the discussion above. They will have a $%
C^{2} $ or $C^{1}$ phase space. The ground state droplet will have $SO(3)$
symmetry in $R^{4}$ bounded by an $S^{3}~$or $U(1)$ symmetry in $R^{2}~$%
bounded by an $S^{1}.$ In the electrostatic picture, there is an $S^{1}$
isometry from the $S^{5},$ that could combine with $\rho $ into a real two
dimensional disk. If we had all the 1/2 BPS geometries for these theories,
they should have a $S^{3}~$isometry out of $S^{5}.$ The ground states have
an enhanced symmetry of $S^{1}~$which combines $\rho ~$to form a disk. Small
ripples on the edge of the disk then describe BPS particles traveling along
the $S^{1}~$equator of the $S^{5}.$ We have a further evidence from the 3+1
d $U(N)$ $\mathcal{N}=4$ SYM on $R\times S^{3}/Z_{k}$ with $k/N$ small. In
the electrostatic picture, the vacuum are periodic disks. If we look at the
vacuum with one disk in a single period, because $k/N$ is small, they look
like charges in a cylinder, and the Laplace equation will be two dimensional
in the space of $\rho $ and an $S^{1}.$ On the disk, the charges feel
logarithmic repulsions plus an spherical symmetric quadratic potential. The
integral equation for the equilibrium configuration is exactly the same as
the one in eqn (2.16) of \cite{Berenstein:2005aa} for the 1/2 BPS ground
state droplet of $\mathcal{N}=4$ SYM on $R\times S^{3}$. We see that their
descriptions are consistent and should be exactly the same for $k=1$.

These ideas should be compared with the gravity results. A class of 1/4 BPS
and 1/8 BPS geometries in $\mathcal{N}=4$ SYM, relevant to the matrix model
discussed above, have been studied by \cite{Donos:2006iy},\cite{Lunin:2006xr}
and by \cite{Kim:2005ez}. They have $U(1)\times SO(4)\times SO(2)_{R}$ \cite%
{Donos:2006iy},\cite{Lunin:2006xr} and $U(1)\times SO(4)~$symmetries \cite%
{Kim:2005ez} (after analytical continuation of $AdS_{3}$) respectively. A
Kahler structure were observed in both cases. In the 1/4 BPS case, it was
reduced to a five dimensional Monge-Amp\`{e}re type equation which is highly
non-linear. The boundary condition is roughly to divide regions where the $%
S^{3}$ or $S^{1}$ shrinks smoothly. Thereby there could be a droplet
configuration in a four dimensional space. However, the regularity condition
and the topology of these solutions are very complicated. To understand the
differential equation in terms of the emergent forces between eigenvalues is
a very challenging problem. Our case of the emergence of the electric charge
picture of eigenvalues from gauge theories may bear some similarities to
these more complicated cases.

The emergence of the odd dimensional sphere from phase space discussed above
can be generalized to other cases, for example $S^{7}$. We consider the
emergence of $AdS_{4}\times S^{7}$ from M2 brane theory on $R\times S^{2}.$
Apart from the time and $S^{2}$ that are already present, we need to explain
the radial coordinate $r$ in $AdS_{4}$ and the $S^{7}.~$We can look at the
1/16 BPS states of the M2 brane theory, by looking at a BPS bound in which
energy is bounded by the sum of 4 $U(1)$ $R$-charges out of $SO(8)_{R}$.
They correspond to excitations of four oscillators associated with the 4
complex scalars for the 1/16 BPS chiral primary states. The 1/16 BPS sector has
an eight dimensional phase space. The ground state droplet should have an
enhanced symmetry of $S^{7}$ in the phase space$.$ Similar picture for the
1/2 BPS sector of M2 brane theory has been studied in \cite{Lin:2004nb}, the
emergence of the $S^{1}~($part of $S^{7})~$for the ground state in a
two-plane has been observed.

Other ways of emergence of odd dimensional spheres, similar to the even
dimensional ones, have been studied by for example \cite{Guralnik:2000pb},%
\cite{Ramgoolam:2001zx},\cite{Sheikh-Jabbari:2004ik}. Their emergence is
similar to the $S^{2}$ in our cases.

\section{Conclusion and discussions}

\renewcommand{\theequation}{5.\arabic{equation}} \setcounter{equation}{0}

In this paper we studied the vacua of gauge theories and the dual gravity
descriptions of each vacuum. These theories are $U(N)$ plane-wave matrix
model and 2+1 dimensional SYM on $R\times S^{2}$. They have a large number
of supersymmetric vacua. Each vacuum can be consistently lifted to the $%
\mathcal{N}=4$ SYM on $R\times S^{3},$ where they become pure gauge
transformations of the vacuum of the $\mathcal{N}=4$ SYM \cite{5brane}.
There exist instanton solutions interpolating some of these vacua. Explicit
instanton solutions are known. The conditions for allowed instanton
solutions are determined by the initial and final vacua and the gauge group.
In the representation of Young tableaux, the condition becomes moving a
number of boxes from left to right only. The instanton actions are the
difference of superpotentials of initial and final vacua. We compute the
superpotential of each vacuum on the weak coupling gauge theory side, and
they are traces of some operators and their final answers depend on the
coupling and the numbers like the $N_{2}^{(i)}, N_{5}^{(i)}$.

We then study the description of these instantons in gravity duals. In the
case where initial and final vacua are very close to each other, the
instanton can be approximated as an Euclidean brane wrapping certain cycle
in the gravity dual background. Each gravity background dual to the vacuum
is given by the configuration of an electric charge system where the charges
distribute on disks under an external quadratic potential. The action of
such Euclidean D2-brane wrapping a three-cycle shows that it is the energy
difference between initial and final configurations of the electrostatic
system. We thereby are able to know the instanton actions corresponding to
instantons interpolating quite different backgrounds since we know the
superpotentials. We define the superpotential of each vacuum on gravity side
by the energy of the charge system or eigenvalue system. When the
eigenvalues or charges do not interact with each other, they do not form
disks and this agrees with the superpotential in weak coupling gauge theory
analysis. When considering quantum effects of this eigenvalue system, the
charges repel and form extended disks. The non-perturbative superpotential
could be obtained by solving some integral equations. We also analyzed the
case of a Euclidean NS5 brane wrapping a six-cycle, and it shows strong
coupling effect of the superpotential which is not seen on the weak coupling
gauge theory side.

Though the instanton solutions exist, they do not contribute to the
correction of the energies of these vacua. These instantons preserve eight
supercharges and break half of the supersymmetries of the background. There
are fermionic zero modes including the eight Goldstone fermions from broken
supersymmetries and integrations of the fermionic collective coordinates in
the path integral make the instanton tunneling amplitude vanish.

We see that the geometry and the disk configuration on the gravity side is
an emergent phenomenon from the eigenvalues of the matrices in the boundary
gauge theories. On the gravity side, the $S^{2}$ emerges due to commutation
relations of $SO(3)$ scalars. The $\eta$ direction emerges as the
eigenvalues of an adjoint scalar $\Phi$. Further we study the emergence of $%
S^{5}$ and $\rho$ direction qualitatively. The ground state geometries we
have is a subsector of the 1/8 BPS geometries, associated with three $R$%
-charges. The sector of the 1/8 BPS states in these theories can be
described by three dimensional harmonic oscillators associated with the
three complex scalars in these theories. They have a six dimensional phase
space. The $S^{5}$ emerges because it is the enhanced symmetry of the ground
state droplet in the phase space. The charges are extended so there is
another radial direction in the phase space, which give rise to the $\rho$
direction on the gravity side.

We characterize the feature of the emergence of the electrostatic picture
qualitatively. The forces of the charges are from the quantum repulsions of
the eigenvalues. However, we are not able to derive the force
quantitatively, which is a Green's function in three dimensions for the
Laplace operator. In the case of the theory of $U(N)$ $\mathcal{N}=4$ SYM on 
$R\times S^{3}/Z_{k},$ with $k/N$ small, the force becomes a two dimensional
Green's function, and this is almost the same as the $\mathcal{N}=4$ SYM on $%
R\times S^{3}$ previously known. Perhaps a more concrete approach to derive
the force from the gauge theory side is to use a superfield formulation of these theories, and study
the loop equations of the chiral ring operators in these theories, similar
to the method in \cite{Cachazo:2002ry}. The loop equations might give us
some equations that are the gauge theory results of the forces between
eigenvalues.

If we can derive the force, by the definition of the non-perturbative
superpotential in terms of the energy of the eigenvalue system, we will be
able to know the non-perturbative superpotentials from gauge theory side.
They will also reveal the quantum nature of the NS5 branes in these theories.

The theories we studied are similar to the 3+1 dimensional $\mathcal{N}%
=1^{\ast }$ SYM \cite{Vafa:1994tf},\cite{Polchinski:2000uf}. They all have
many vacua and mass gaps. Some of the vacua discribe NS5 branes. Most of the
dual geometries are smooth and similar embeddings of branes are possible if
we know the full solutions dual to $\mathcal{N}=1^{\ast }$ SYM. The features
of the gravity duals of these theories in this paper might help us gain some
information that is similar to those of the $\mathcal{N}=1^{\ast }$ SYM. The
NS5 branes in the $\mathcal{N}=1^{\ast }$ SYM wrap $R^{3,1}\times S^{2},$
while in our cases they wrap $R\times S^{5},$ under the limit when the $S^{2}
$ and $S^{5}~ $become infinitely large, these two theories can be related
via T dualities three times. This limit corresponds to expanding around the
tip of a pair of disks in the plane-wave matrix model. There is a curious
similarity with the eigenvalue distributions in this limit (cf. (4.2) of 
\cite{Dijkgraaf:2002dh}) for the trivial vacuum of $\mathcal{N}=1^{\ast }$
SYM. It would be interesting to understand better the NS5 brane vacua and
the little string theory from these theories.

Similar to the two theories studied in this paper, the vacua of the $%
\mathcal{N}=4$ SYM on $R\times S^{3}/Z_{k}$ \cite{5brane},\cite%
{Ishiki:2006rt} and the IIA NS5 brane theory on $R\times S^{5}$ \cite{5brane}%
,\cite{Ling:2006up} are also characterized by charged disk configurations.
It would be interesting to understand the instanton solutions in these
theories. The disk configurations in these four theories can be related to
each other by various limits.

The plane-wave matrix model is also similar to the tiny graviton matrix
theory in \cite{Sheikh-Jabbari:2004ik}. The model in \cite%
{Sheikh-Jabbari:2004ik} has solutions corresponding to fuzzy three-spheres
and different vacua correspond to different representations of $Spin(4)$. It
is possible that there are also similar instanton solutions in \cite%
{Sheikh-Jabbari:2004ik}. This may help in understanding the instantons
interpolating between different giant D3-branes \cite{Hashimoto:2000zp},\cite%
{Grisaru:2000zn}.

\section*{Acknowledgments}

I am very grateful to Juan Maldacena for many important discussions. I also
thank I. Bena, R. Bousso, S. Cremonini, N. Itzhaki, F. Larsen, J. T. Liu, O.
Lunin, L. Pando Zayas, L. Rastelli, J. Shao, P. Shepard, M. Spradlin, D.
Vaman for helpful conversations. The research was supported in part by NSF
Grant No. PHY-0243680 (Princeton University), DOE grant \#DE-FG02-90ER40542
(IAS, Princeton) and DOE grant \#DE-FG02-95ER40899 (MCTP). I also thank the
hospitality of U.C. Berkeley where some of this work was done.


\begin{thebibliography}{99}
\bibitem{Maldacena:1997re} J.~M.~Maldacena, ``The large N limit of
superconformal field theories and supergravity,'' Adv.\ Theor.\ Math.\
Phys.\ \textbf{2}, 231 (1998) [Int.\ J.\ Theor.\ Phys.\ \textbf{38}, 1113
(1999)] [arXiv:hep-th/9711200]. 

\bibitem{Gubser:1998bc} S.~S.~Gubser, I.~R.~Klebanov and A.~M.~Polyakov,
``Gauge theory correlators from non-critical string theory,'' Phys.\ Lett.\
B \textbf{428}, 105 (1998) [arXiv:hep-th/9802109]. 

\bibitem{Witten:1998qj} E.~Witten, ``Anti-de Sitter space and holography,''
Adv.\ Theor.\ Math.\ Phys.\ \textbf{2}, 253 (1998) [arXiv:hep-th/9802150]. 

\bibitem{Berenstein:2005aa} D.~Berenstein, ``Large N BPS states and emergent
quantum gravity,'' JHEP \textbf{0601}, 125 (2006) [arXiv:hep-th/0507203]. 

\bibitem{Corley:2001zk} S.~Corley, A.~Jevicki and S.~Ramgoolam, ``Exact
correlators of giant gravitons from dual N = 4 SYM theory,'' Adv.\ Theor.\
Math.\ Phys.\ \textbf{5}, 809 (2002) [arXiv:hep-th/0111222]. 

\bibitem{Berenstein:2004kk} D.~Berenstein, ``A toy model for the AdS/CFT
correspondence,'' JHEP \textbf{0407}, 018 (2004) [arXiv:hep-th/0403110]. 

\bibitem{Lin:2004nb} H.~Lin, O.~Lunin and J.~M.~Maldacena, ``Bubbling AdS
space and 1/2 BPS geometries,'' JHEP \textbf{0410}, 025 (2004)
[arXiv:hep-th/0409174]. 

\bibitem{Berenstein:2002jq} D.~Berenstein, J.~M.~Maldacena and H.~Nastase,
``Strings in flat space and pp waves from N = 4 super Yang Mills,'' JHEP 
\textbf{0204}, 013 (2002) [arXiv:hep-th/0202021]. 

\bibitem{Maldacena:2002rb} J.~M.~Maldacena, M.~M.~Sheikh-Jabbari and M.~Van
Raamsdonk, ``Transverse fivebranes in matrix theory,'' JHEP \textbf{0301},
038 (2003) [arXiv:hep-th/0211139]. 

\bibitem{5brane} H.~Lin and J.~Maldacena, ``Fivebranes from gauge theory,''
arXiv:hep-th/0509235. 

\bibitem{Dijkgraaf:2002dh} R.~Dijkgraaf and C.~Vafa, ``A perturbative window
into non-perturbative physics,'' arXiv:hep-th/0208048. 

\bibitem{Dasgupta:2002hx} K.~Dasgupta, M.~M.~Sheikh-Jabbari and M.~Van
Raamsdonk, ``Matrix perturbation theory for M-theory on a PP-wave,'' JHEP 
\textbf{0205}, 056 (2002) [arXiv:hep-th/0205185]. 

\bibitem{Kim:2002if} N.~Kim and J.~Plefka, \textquotedblleft On the spectrum
of pp-wave matrix theory,\textquotedblright\ Nucl.\ Phys.\ B \textbf{643},
31 (2002) [arXiv:hep-th/0207034]. 

\bibitem{Dasgupta:2002ru} K.~Dasgupta, M.~M.~Sheikh-Jabbari and M.~Van
Raamsdonk, ``Protected multiplets of M-theory on a plane wave,'' JHEP 
\textbf{0209}, 021 (2002) [arXiv:hep-th/0207050]. 

\bibitem{Kim:2002zg} N.~Kim and J.~H.~Park, ``Superalgebra for M-theory on a
pp-wave,'' Phys.\ Rev.\ D \textbf{66}, 106007 (2002) [arXiv:hep-th/0207061]. 

\bibitem{Witten:1981nf} E.~Witten, ``Dynamical Breaking Of Supersymmetry,''
Nucl.\ Phys.\ B \textbf{188}, 513 (1981). 

\bibitem{Salomonson:1981ug} P.~Salomonson and J.~W.~van Holten, ``Fermionic
Coordinates And Supersymmetry In Quantum Mechanics,'' Nucl.\ Phys.\ B 
\textbf{196}, 509 (1982). 

\bibitem{Ling:2006up} H.~Ling, A.~R.~Mohazab, H.~H.~Shieh, G.~van Anders and
M.~Van Raamsdonk, ``Little string theory from a double-scaled matrix
model,'' arXiv:hep-th/0606014. 

\bibitem{Ebrahim:2005sb} H.~Ebrahim, ``Semiclassical strings probing NS5
brane wrapped on S**5,'' JHEP \textbf{0601}, 019 (2006)
[arXiv:hep-th/0511228]. 

\bibitem{Park:2002cb}  J.~H.~Park,  ``Supersymmetric objects in the M-theory
on a pp-wave,''  JHEP \textbf{0210}, 032 (2002)  [arXiv:hep-th/0208161].  

\bibitem{Yee:2003ge} J.~T.~Yee and P.~Yi, ``Instantons of M(atrix) theory in
pp-wave background,'' JHEP \textbf{0302}, 040 (2003) [arXiv:hep-th/0301120]. 

\bibitem{Lee:2005dd} H.~K.~Lee, \textquotedblleft Gauge theory and
supergravity duality in the pp-wave background,\textquotedblright\ Ph.D.
thesis, Caltech, 2005, UMI-31-97350. 

\bibitem{Bachas:2000dx} C.~Bachas, J.~Hoppe and B.~Pioline, ``Nahm
equations, N = 1* domain walls, and D-strings in AdS(5) x S(5),'' JHEP 
\textbf{0107}, 041 (2001) [arXiv:hep-th/0007067]. 

\bibitem{Jatkar:2001uh}  D.~P.~Jatkar, G.~Mandal, S.~R.~Wadia and
K.~P.~Yogendran,  ``Matrix dynamics of fuzzy spheres,''  JHEP \textbf{0201},
039 (2002)  [arXiv:hep-th/0110172].  

\bibitem{Vafa:1994tf} C.~Vafa and E.~Witten, ``A Strong coupling test of S
duality,'' Nucl.\ Phys.\ B \textbf{431}, 3 (1994) [arXiv:hep-th/9408074]. 

\bibitem{Polchinski:2000uf} J.~Polchinski and M.~J.~Strassler,
\textquotedblleft The string dual of a confining four-dimensional gauge
theory,\textquotedblright\ arXiv:hep-th/0003136. 

\bibitem{Ishiki:2006rt} G.~Ishiki, Y.~Takayama and A.~Tsuchiya, ``N = 4 SYM
on R x S**3 and theories with 16 supercharges,'' arXiv:hep-th/0605163. 

\bibitem{Kim:2003rz}  N.~w.~Kim, T.~Klose and J.~Plefka,  ``Plane-wave
matrix theory from N = 4 super Yang-Mills on R x S**3,''  Nucl.\ Phys.\ B 
\textbf{671}, 359 (2003)  [arXiv:hep-th/0306054].  

\bibitem{Klose:2003qc}  T.~Klose and J.~Plefka, ``On the integrability of
large N plane-wave matrix theory,''  Nucl.\ Phys.\ B \textbf{679}, 127
(2004)  [arXiv:hep-th/0310232].  

\bibitem{Fischbacher:2004iu}  T.~Fischbacher, T.~Klose and J.~Plefka, 
``Planar plane-wave matrix theory at the four loop order: Integrability 
without BMN scaling,''  JHEP \textbf{0502}, 039 (2005) 
[arXiv:hep-th/0412331].  

\bibitem{Okuyama:2002zn}  K.~Okuyama,  ``N = 4 SYM on R x S(3) and
pp-wave,''  JHEP \textbf{0211}, 043 (2002)  [arXiv:hep-th/0207067].  

\bibitem{Kabat:1997im} D.~Kabat and W.~I.~Taylor, ``Spherical membranes in
matrix theory,'' Adv.\ Theor.\ Math.\ Phys.\ \textbf{2}, 181 (1998)
[arXiv:hep-th/9711078]. 

\bibitem{Maldacena:2001xj} J.~M.~Maldacena, G.~W.~Moore and N.~Seiberg,
``D-brane instantons and K-theory charges,'' JHEP \textbf{0111}, 062 (2001)
[arXiv:hep-th/0108100]. 

\bibitem{McGreevy:2000cw}  J.~McGreevy, L.~Susskind and N.~Toumbas, 
``Invasion of the giant gravitons from anti-de Sitter space,''  JHEP \textbf{%
0006}, 008 (2000)  [arXiv:hep-th/0003075].  

\bibitem{Hashimoto:2000zp} A.~Hashimoto, S.~Hirano and N.~Itzhaki, ``Large
branes in AdS and their field theory dual,'' JHEP \textbf{0008}, 051 (2000)
[arXiv:hep-th/0008016]. 

\bibitem{Grisaru:2000zn} M.~T.~Grisaru, R.~C.~Myers and O.~Tafjord, ``SUSY
and Goliath,'' JHEP \textbf{0008}, 040 (2000) [arXiv:hep-th/0008015]. 

\bibitem{Das:2000fu} S.~R.~Das, A.~Jevicki and S.~D.~Mathur, ``Giant
gravitons, BPS bounds and noncommutativity,'' Phys.\ Rev.\ D \textbf{63},
044001 (2001) [arXiv:hep-th/0008088]. 

\bibitem{Bandos:2000az} I.~A.~Bandos, A.~Nurmagambetov and D.~P.~Sorokin,
``The type IIA NS5-brane,'' Nucl.\ Phys.\ B \textbf{586}, 315 (2000)
[arXiv:hep-th/0003169]. 

\bibitem{Papadopoulos:1997je} G.~Papadopoulos, ``T-duality and the
worldvolume solitons of five-branes and KK-monopoles,'' Phys.\ Lett.\ B 
\textbf{434}, 277 (1998) [arXiv:hep-th/9712162]. 

\bibitem{Bena:2000fz} I.~Bena and A.~Nudelman, ``Warping and vacua of
(S)YM(2+1),'' Phys.\ Rev.\ D \textbf{62}, 086008 (2000)
[arXiv:hep-th/0005163]. 

\bibitem{Bena:2000va} I.~Bena and A.~Nudelman, ``Exotic polarizations of D2
branes and oblique vacua of (S)YM(2+1),'' Phys.\ Rev.\ D \textbf{62}, 126007
(2000) [arXiv:hep-th/0006102]. 

\bibitem{Cachazo:2001jy} F.~Cachazo, K.~A.~Intriligator and C.~Vafa, ``A
large N duality via a geometric transition,'' Nucl.\ Phys.\ B \textbf{603},
3 (2001) [arXiv:hep-th/0103067]. 

\bibitem{Itzhaki:1998dd} N.~Itzhaki, J.~M.~Maldacena, J.~Sonnenschein and
S.~Yankielowicz, ``Supergravity and the large N limit of theories with
sixteen supercharges,'' Phys.\ Rev.\ D \textbf{58}, 046004 (1998)
[arXiv:hep-th/9802042]. 

\bibitem{Lin:2004kw} H.~Lin, ``The supergravity dual of the BMN matrix
model,'' JHEP \textbf{0412}, 001 (2004) [arXiv:hep-th/0407250]. 

\bibitem{Nastase:2004te} H.~Nastase, ``On fuzzy spheres and (M)atrix
actions,'' arXiv:hep-th/0410137. 

\bibitem{Lozano:2005kf} Y.~Lozano and D.~Rodriguez-Gomez, ``Fuzzy 5-spheres
and pp-wave Matrix actions,'' JHEP \textbf{0508}, 044 (2005)
[arXiv:hep-th/0505073]. 

\bibitem{Taylor:1999gq} W.~I.~Taylor and M.~Van Raamsdonk, ``Multiple
D0-branes in weakly curved backgrounds,'' Nucl.\ Phys.\ B \textbf{558}, 63
(1999) [arXiv:hep-th/9904095]. 

\bibitem{love} E. R. Love, The electrostatic field of two equal circular
co-axial conducting disks, The Quarterly Journal of Mechanics and Applied
Mathematics, 1949 2(4):428-451.

\bibitem{kp} M. Kac, H. Pollard, The distribution of the maximum of partial
sums of independent random variables, Canad. J. Math. 2 (1950), 375-384.

\bibitem{hutson} V. Hutson, The circular plate condenser at small
separations, Proc. Cambridge Phil. Soc. 59, 211-224 (1963).

\bibitem{Castelino:1997rv} J.~Castelino, S.~M.~Lee and W.~I.~Taylor,
``Longitudinal 5-branes as 4-spheres in matrix theory,'' Nucl.\ Phys.\ B 
\textbf{526}, 334 (1998) [arXiv:hep-th/9712105]. 

\bibitem{Ramgoolam:2001zx} S.~Ramgoolam, ``On spherical harmonics for fuzzy
spheres in diverse dimensions,'' Nucl.\ Phys.\ B \textbf{610}, 461 (2001)
[arXiv:hep-th/0105006]. 

\bibitem{deMelloKoch:2002nq}  R.~de Mello Koch, A.~Jevicki and
J.~P.~Rodrigues,  ``Collective string field theory of matrix models in the
BMN limit,''  Int.\ J.\ Mod.\ Phys.\ A \textbf{19}, 1747 (2004) 
[arXiv:hep-th/0209155].  

\bibitem{Berenstein:2005jq}  D.~Berenstein, D.~H.~Correa and S.~E.~Vazquez, 
``All loop BMN state energies from matrices,''  JHEP \textbf{0602}, 048
(2006)  [arXiv:hep-th/0509015].  

\bibitem{Donos:2005vm} A.~Donos, A.~Jevicki and J.~P.~Rodrigues, ``Matrix
model maps in AdS/CFT,'' arXiv:hep-th/0507124. 

\bibitem{Rodrigues:2005ec}  J.~P.~Rodrigues,  ``Large N spectrum of two
matrices in a harmonic potential and BMN  energies,''  JHEP \textbf{0512},
043 (2005)  [arXiv:hep-th/0510244].  

\bibitem{Vazquez:2006hd}  S.~E.~Vazquez,  ``BPS condensates, matrix models
and emergent string theory,''  arXiv:hep-th/0607204.  

\bibitem{Caldarelli:2004ig} M.~M.~Caldarelli and P.~J.~Silva, ``Giant
gravitons in AdS/CFT. I: Matrix model and back reaction,'' JHEP \textbf{0408}%
, 029 (2004) [arXiv:hep-th/0406096]. 

\bibitem{Takayama:2005yq} Y.~Takayama and A.~Tsuchiya, ``Complex matrix
model and fermion phase space for bubbling AdS geometries,'' JHEP \textbf{%
0510}, 004 (2005) [arXiv:hep-th/0507070]. 

\bibitem{Yoneya:2005si} T.~Yoneya, ``Extended fermion representation of
multi-charge 1/2-BPS operators in AdS/CFT: Towards field theory of
D-branes,'' JHEP \textbf{0512}, 028 (2005) [arXiv:hep-th/0510114]. 

\bibitem{Donos:2006iy} A.~Donos, ``A description of 1/4 BPS configurations
in minimal type IIB SUGRA,'' arXiv:hep-th/0606199. 

\bibitem{Lunin:2006xr} O. Lunin, unpublished notes, 2004; see also,
O.~Lunin, \textquotedblleft On gravitational description of Wilson
lines,\textquotedblright\ JHEP \textbf{0606}, 026 (2006)
[arXiv:hep-th/0604133]. 

\bibitem{Kim:2005ez} N.~Kim, ``AdS(3) solutions of IIB supergravity from
D3-branes,'' JHEP \textbf{0601}, 094 (2006) [arXiv:hep-th/0511029]. 

\bibitem{Guralnik:2000pb} Z.~Guralnik and S.~Ramgoolam, ``On the
polarization of unstable D0-branes into non-commutative odd spheres,'' JHEP 
\textbf{0102}, 032 (2001) [arXiv:hep-th/0101001]. 

\bibitem{Sheikh-Jabbari:2004ik} M.~M.~Sheikh-Jabbari, ``Tiny graviton matrix
theory: DLCQ of IIB plane-wave string theory, a conjecture,'' JHEP \textbf{%
0409}, 017 (2004) [arXiv:hep-th/0406214]. 

\bibitem{Cachazo:2002ry} F.~Cachazo, M.~R.~Douglas, N.~Seiberg and
E.~Witten, ``Chiral rings and anomalies in supersymmetric gauge theory,''
JHEP \textbf{0212}, 071 (2002) [arXiv:hep-th/0211170]. 
\end{thebibliography}
\end{document}